# The human brain's network architecture is genetically encoded by modular pleiotropy

Maxwell A. Bertolero[1], Ann Sizemore Blevins[1], Graham L. Baum[2], Ruben C. Gur[2], Raquel E. Gur[2], David R. Roalf[2], Theodore D. Satterthwaite[2], Danielle S. Bassett[1-7]

[1]Department of Bioengineering, School of Engineering & Applied Science, University of Pennsylvania, Philadelphia, PA 19104 USA
[2]Department of Psychiatry, Perelman School of Medicine, University of Pennsylvania, Philadelphia, PA 19104 USA
[3]Department of Electrical & Systems Engineering, School of Engineering & Applied Science, University of Pennsylvania, Philadelphia, PA, 19104 USA
[4]Department of Neurology, Perelman School of Medicine, University of Pennsylvania, Philadelphia, PA, 19104 USA
[5]Department of Physics & Astronomy, College of Arts & Sciences, University of Pennsylvania, Philadelphia, PA, 19104 USA
[6]Santa Fe Institute, Santa Fe, NM, 87501 USA

[7]To whom correspondence should be address: dsb@seas.upenn.edu

For much of biology, the manner in which genotype maps to phenotype remains a fundamental mystery. The few maps that are known tend to show *modular pleiotropy*: sets of phenotypes are determined by distinct sets of genes. One key map that has evaded discovery is that of the human brain's network architecture. Here, we determine the form of this map for gene coexpression and single nucleotide polymorphisms. We discover that mostly non-overlapping sets of genes encode the connectivity of brain network modules (or so-called *communities*), suggesting that brain network communities demarcate genetic transitions. We find that these clean boundaries break down at connector hubs, whose integrative connectivity is encoded by pleiotropic genes from mostly non-overlapping sets. Broadly, this study opens fundamentally new directions in the study of genetic encoding of brain development, evolution, and disease.

Many complex biological systems—from the musculoskeletal system to protein interactions—can be parsimoniously represented as networks composed of elements (nodes) and their interactions or relations (edges). As a quintessential example, the human brain is composed of large areas interlinked by structural or functional connections. The coarse-grained organization of the brain's network architecture is modular, where groups of nodes tend to form tightly interconnected networks or "communities"[1,2]. Each community is dedicated to a specific class of cognitive functions, and, collectively, communities are integrated by regions known as connector hubs that have connections equally spread across the network's communities[3–13]. Communities and connector hubs are features of brain network architecture that are expressed consistently but also vary among individuals[14,15] in a manner that predicts cognitive performance[16–18], tracks response to cognitive training[19], differentiates mental disorders, guides effective treatments[20,21], and changes over development[22] in concert with changes in myelination and cortical thickness[23]. Finally, brain network organization as manifest in both structural[24] and functional connectivity[25–27] is heritable, suggesting that brain connectivity is genetically encoded. However, the form of this encoding is unknown.

Genotype-phenotype relationships typically exhibit modular pleiotropy—sets of phenotypes are determined by mostly non-overlapping sets of genes[28,29]. It is interesting to consider the possibility that the map of gene-brain relations could be parsimoniously explained by modular pleiotropy[2,4,11]. In other words, genes could be organized into mostly non-overlapping sets, where each set encodes the connectivity of a single brain network community. Specifically, the coexpression of a particular gene set could statistically explain the connectivity of one brain community but not the connectivity of any other brain community[30–33]. Moreover, single nucleotide polymorphisms (SNPs) at particular genes could alter the connectivity of one brain community but not the connectivity of any other brain community. If such a mapping existed, it would suggest that brain communities comprise bounded areas that demarcate genetic transitions, and therefore that gene encoding should change not as one moves across the cortex but as one crosses a community border. Moreover, while gene expression and SNPs could both be encoded via modular pleiotropy, they might play different roles in the brain's network architecture, its evolution and development, and its heritability and conservation.

Here, we address this possibility in a multimodal study combining gene expression from six post-mortem brains from the Allen Brain Institute, functional and structural brain connectivity, genotyping, and cognitive and behavioral testing of 895 young adults in the Human Connectome Project (S1200), functional and structural brain connectivity and neuropsychological test scores from 380 youth (ages 8-13 years) in the Philadelphia Neurodevelopmental Cohort, and previously published evolutionary expansion data from macaques to humans[34]. First, we determine which genes' coexpression and SNPs capture each brain region's functional and structural connectivity. We find that for each community in the brain's network, mostly non-overlapping sets of genes' coexpression and SNPs capture that community's connectivity. Second, we find meaningful variance in how poorly or how well gene coexpression and SNPs capture each brain region's connectivity. Gene coexpression and SNPs best capture different brain region's connectivity. This difference relates to each brain regions' network role, each region's contribution to performance during cognitive tasks, the development of brain connectivity and executive function, brain heritability, brain conservation, and the evolutionary expansion of the cortex from macaques to humans. In sum, we discover a modular pleiotropic genetic map of the human brain's network architecture, and demonstrate that SNPs and gene coexpression differentially relate to brain evolution, heritability, network architecture, conservation, and development.

# Gene coexpression maps preferentially to local functional connectivity

We first sought to measure the extent to which gene coexpression could statistically account for patterns of functional and structural connectivity. To quantitatively probe the relation between gene coexpression and brain connectivity, we first defined a metric that we refer to as *gene coexpression fit*: a brain region's gene coexpression fit is given by the Pearson correlation coefficient $r$ between that region's gene coexpression with all other regions and that region's connectivity strength (functional or structural) with all other regions after accounting for physical distance (see Methods)[35]. Next, we developed a machine-learning algorithm to find the set of genes that maximized the gene coexpression fit for each region using simulated annealing (see Methods for details on the algorithm, the number of genes used, and the comparison to null models; **Fig. 1**, **Extended Data Fig. 1, Extended Data Fig. 2, Extended Data Fig. 3, Extended Data Fig. 4, Extended Data Fig. 5, Extended Data Fig. 6, Extended Data Fig. 7**). As an initial test of biological plausibility, we examined whether the genes involved in fitting structural and functional connectivity were also involved in processes that are relevant for neural function using a gene ontology enrichment analysis (GOrilla[36]; see methods). Critically, for both structural and functional connectivity, many gene ontologies relevant to brain function and structure were found to be significantly enriched after correcting for multiple comparisons (See Methods, **Extended Data Fig. 8, Extended Data Fig. 9)**.

We next sought to determine which types of nodes in the network could be best captured by gene coexpression. We modeled functional or structural connectivity as a network in which network nodes represent brain regions and a network edge represented a functional or structural connection between two nodes. Critically, calculating the gene coexpression fit for each region allowed us to determine whether gene coexpression fits differed across regions that play distinct roles within the network. We focused on two specific roles that both relate to the diversity of a node's edges across communities in the network[2,37]. First, we considered local nodes, which we defined as nodes with low participation coefficients, indicating that they do not tend to link communities together. Second, we considered connector hubs, which we defined as nodes with high participation coefficients, indicating that they tend to link many communities together[4–6,38] (see Methods). We found that gene coexpression fits were negatively correlated with the participation coefficient, demonstrating that gene coexpression fits were higher for local nodes than for connector nodes (**Fig. 1**; **Extended Data Fig. 10**).

We also found that gene coexpression fits were higher for functional networks than for structural networks (**Fig. 1, Extended Data Fig. 10**). To determine the anatomical specificity of this effect, we averaged the gene coexpression fits of all regions within each network community. We found that gene coexpression fits for functional connectivity were significantly higher than for structural connectivity in 6 of the 7 communities (**Fig. 1**). However, the gene coexpression fit of functional connectivity was positively correlated with the gene coexpression fit of structural connectivity across regions ($r = 0.322$, $-\log_{10}(p)=5$, dof=191), potentially indicating some cross-modality conservation. We then visualized, for each node, which genes' coexpression fit that node's connectivity. While pleiotropic genes exists, the expression of most genes appeared to primarily fit the connectivity of nodes in a single brain community (**Fig. 1c**). This hypothesis is quantitatively tested below.

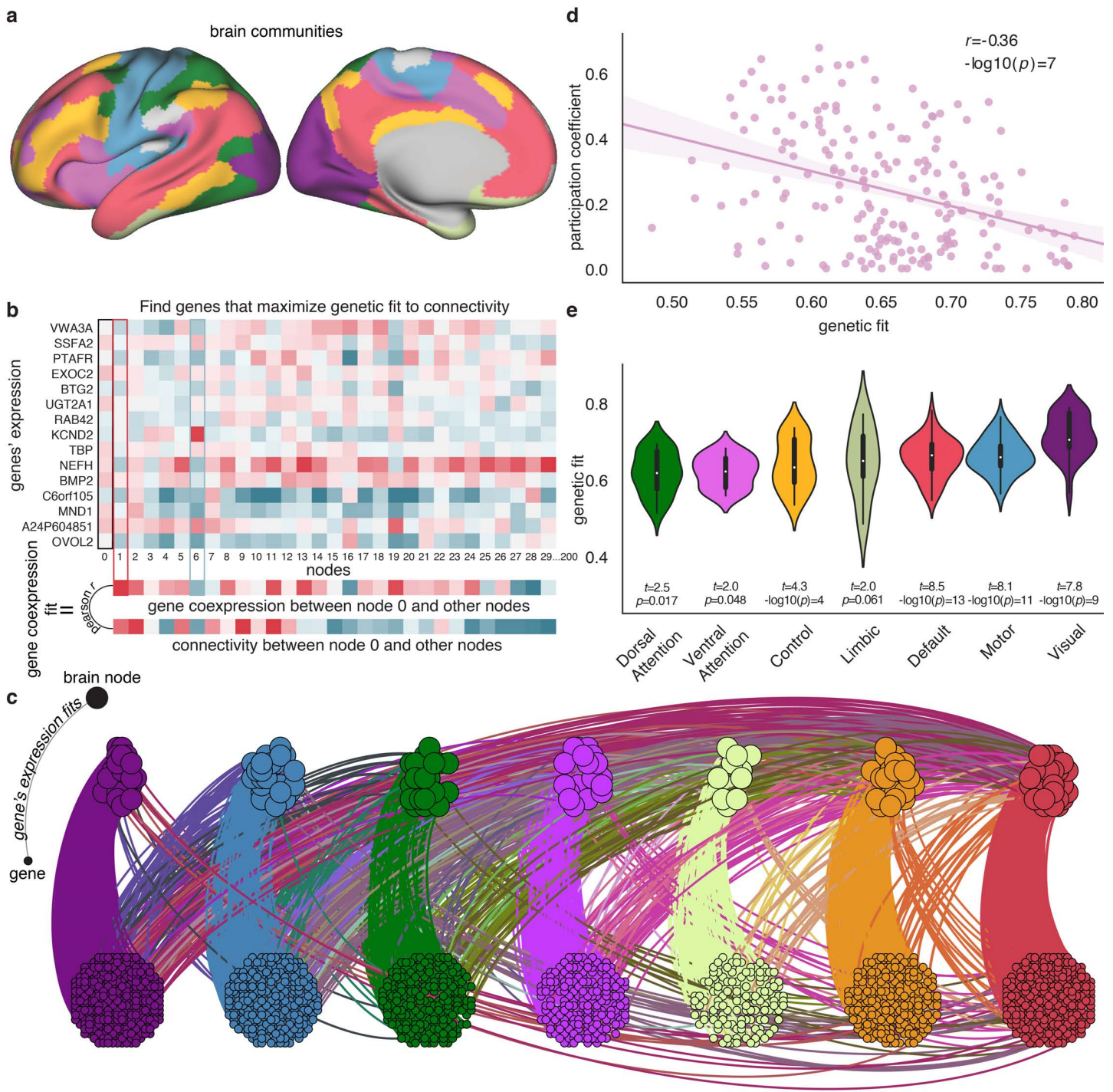

**Fig. 1 | Mapping gene coexpression to brain network connectivity. a,** We used an *a priori* assignment of nodes to communities. **b,** For each node, genes were chosen to maximize the gene coexpression fit, defined as the Pearson correlation coefficient *r* between region 0's gene coexpression with all other regions and region 0's (functional or structural) connectivity strength with all other regions after accounting for interregional Euclidean distance. **c,** An existing edge between two nodes represents the fact that a gene (a node in the bottom layer of the network) was found to fit the connectivity of a brain region (a node in the top layer of the network). Brain nodes are colored by functional community. Each gene is the color of the brain community that the gene was found to most frequently fit. While pleiotropy exists, most genes primarily fit the connectivity of nodes in a single brain community; this hypothesis is quantitatively tested in **Fig. 3**. **d,** The correlation between a region's gene coexpression fit for functional connectivity and the functional participation coefficient (dof=191). For results for separate *n* gene sets and structural connectivity, see **Extended Data Fig. 10**. **e,** The mean gene coexpression fit for regions in each network community. For every community except the Limbic system, the gene coexpression fit was significantly higher for functional connectivity than for structural connectivity (dof=191).

## Single nucleotide polymorphisms at genes whose coexpression captures brain connectivity explain the variance in connectivity across individuals

It is intuitively plausible that SNPs at genes whose coexpression fits a region's connectivity are genes that are responsible for encoding that region's connectivity in the genome. To assess this intuition, we measured how much each gene's variance (in the form of SNPs at that gene) could explain the variance in each region's connectivity across subjects. For each brain region, a GWAS analysis of the region's connectivity (i.e., edges) was executed in PLINK [39][40] (see Methods). For each gene and brain region, the mean absolute linear regression coefficient (beta) across SNPs at that gene was calculated. Next, the mean beta of edges at that brain region was calculated. This process generates a region by gene array representing how well variance (SNPs) at each gene explains the inter-individual variance in network connectivity at each region. In what follows, we refer to these beta values as *SNP fits*. The entire process was separately performed for functional connectivity and for structural connectivity.

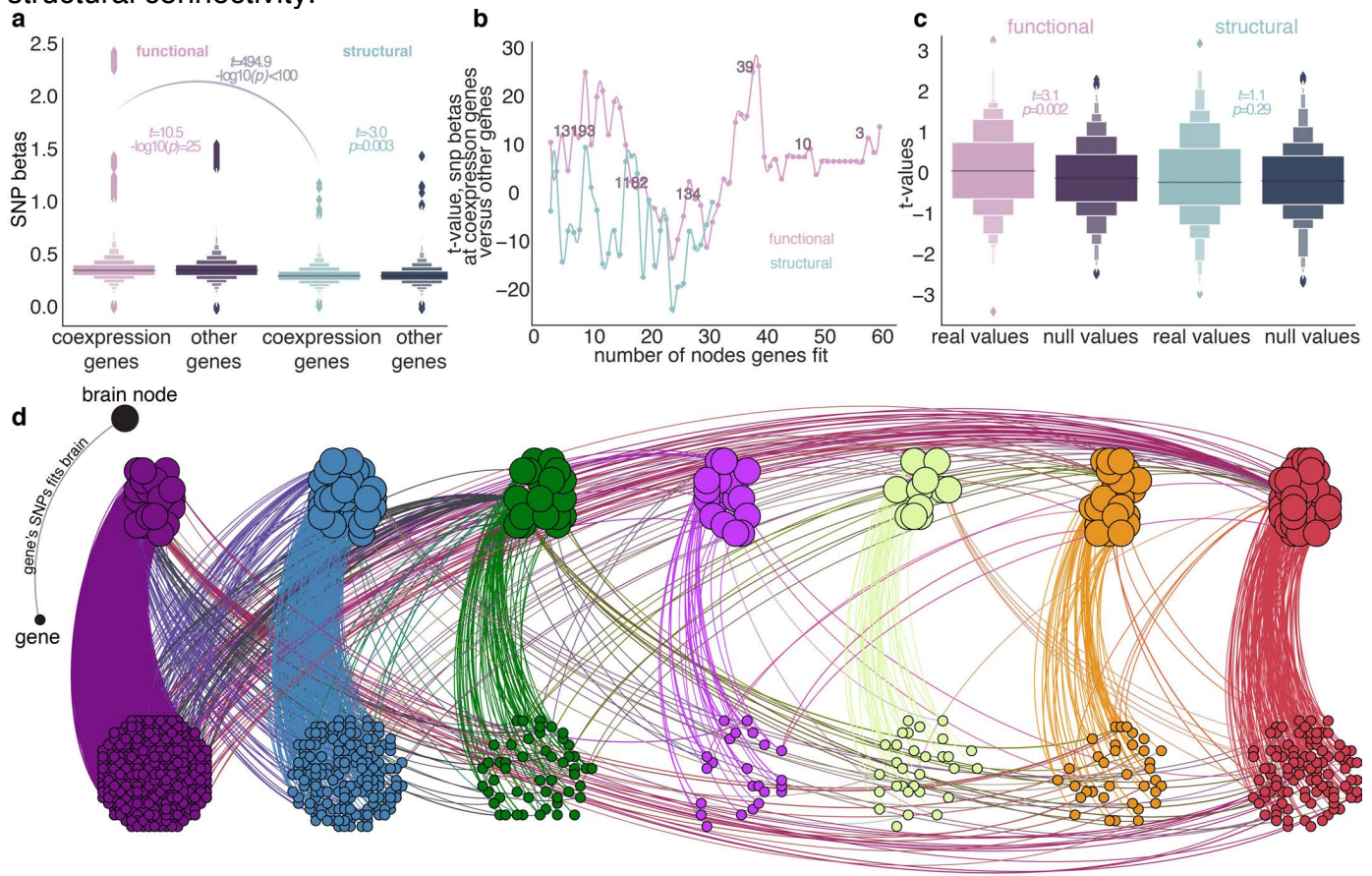

**Fig. 2 | SNPs located at genes whose coexpression fits connectivity tend to explain connectivity variance across subjects**. **a**, We sought to determine whether the SNPs that are located at genes whose coexpression fits brain connectivity are also the SNPs that explain the variance in connectivity across subjects. We therefore compared the distribution of absolute betas for SNPs located at genes whose coexpression frequently ($n=7$, the median) fit brain connectivity to the distribution of SNPs at other genes, separately for both functional and structural connectivity (dof=2669998; see Methods for description of "letter value" boxenplots). **b**, The inclusion threshold can be varied to only include regions that fit a larger number of genes. The number of genes included at each threshold is plotted inline, and the *t*-value (i.e., the *t*-value from panel **a**, but with a different threshold) at each threshold is shown (dof=2669998). In general, SNPs located at genes whose coexpression fit brain connectivity explain the variance in brain connectivity across subjects with significantly greater accuracy than SNPs at genes whose coexpression did not fit brain connectivity. **c**, For each brain region, the *t*-value between SNP betas at genes that fit that region's connectivity and SNP betas at genes that did not fit that region's connectivity was calculated and compared to a distribution of null *t*-values estimated from random divisions of SNP betas. In general, for each brain region, SNPs located at genes whose coexpression fit that region's functional connectivity tended to predict variance at that region's functional connectivity with significantly higher accuracy than a randomly chosen set of genes' SNPs (dof=192). **d**, A visualization of our hypothesis that distinct sets of genes encode the functional connectivity of each brain community. The existence of an edge represents the fact that SNPs located at that gene (a node in the bottom layer of the network) predict the inter-subject variance in connectivity of a brain region (a node in the top layer of the network). Brain regions are colored by their community. Each gene node is the color of the brain community whose connectivity the gene was found to most frequently predict. While pleiotropy exists, most genes primarily predict the connectivity variance of regions in a single functional community. This hypothesis is quantitatively tested in **Fig. 3**.

To determine whether SNPs at genes who coexpression fit brain connectivity also explain brain connectivity variance, we compared the distribution of SNP fits located at genes whose coexpression frequently fit brain connectivity to the distribution of SNP fits located at other genes. For both structural and functional connectivity, the SNP fits at genes whose coexpression frequently fit brain connectivity were higher than the SNP fits at genes whose coexpression did not frequently fit brain connectivity (**Fig. 2a,b**). For functional connectivity, this relationship existed at the regional level as well; in general, a region's functional (but not structural) connectivity variance was better explained by SNPs at genes whose coexpression fit that single region's connectivity than by SNPs at genes whose coexpression did not fit that single region's connectivity (**Fig. 2c**). Finally, SNPs were able to explain significantly more variance in functional connectivity than in structural connectivity ($t=494$, $-\log10(p)<5$, dof=2669998). Similar to gene coexpression, we produced a visualization of the genes whose genetic variance explained the inter-subject variance in a given community's connectivity; the visualization again suggests a modular pleiotropic relationship, in which SNPs at a given gene explain inter-subject variance in the connectivity of regions located primarily in one brain network community (**Fig. 2d**).

## The genetic map of the human brain's network architecture

We next sought to determine whether community boundaries are genetically encoded via modular pleiotropy. We first built a gene coexpression encoding network, where each node is a brain region. An edge between two nodes is determined by the Pearson correlation coefficient *r* between (i) the number of times each gene significantly accounts for the connectivity of the one node and (ii) the number of times each gene significantly accounts for the connectivity of the other node. We also built a SNP encoding network, where each node is a brain region. An edge between two nodes is determined by the Pearson correlation coefficient *r* between (i) the SNP fit values of one node across genes and (ii) the SNP fit values of the other node across genes. Thus, SNP encoding network reflects the degree to which SNPs at the same genes explain variance in connectivity. The gene coexpression encoding network and the SNP encoding network were both constructed separately for functional connectivity and for structural connectivity (**Fig. 3a,b**).

To test for modularity, we then assign each brain region (node) to the canonical communities[2] shown in **Fig. 1a**; we used the 17 community partition due to its detailed boundaries. We then applied this partition to the gene coexpression encoding network and the SNP encoding network, to determine whether gene encoding of brain connectivity follows the boundaries of the canonical communities. If communities represent genetic boundaries, then we expect to obtain a *Q* value from this partitioning that is greater than the *Q* value obtained from appropriate spatial null models (see Methods). Across null models and a range of methodological choices (see Methods), we found that the *Q* values of the real community partitions were significantly higher than the *Q* values of the null community partitions, for the gene coexpression encoding network and the SNP encoding network derived from functional connectivity (**Fig. 3c,d**, one-sample *t*-test, across gene encoding network thresholds of 5,10, and 15 percent functional connectivity $582>t<1676$, $-\log10(p)<100$, dof=9999). We also observed that the edge weights of the gene coexpression encoding network were significantly correlated with the edge weights of the SNP encoding network, when both were derived from functional connectivity (**Fig. 3e**). Intuitively, this finding indicates that if the functional connectivity of two regions tends to be explained by the coexpression of the same set of genes, then the inter-subject variance in the functional connectivity of those two regions tends to be explained by SNPs at the same genes. These relationships did not exist when the gene coexpression encoding network and the SNP encoding network were derived from structural connectivity.

Despite the presence of modularity, the complementary existence of inter-module edges is indicative of pleiotropy. To determine where genetic pleiotropy is highest, we assessed the role that each brain region plays in the gene encoding network. Our intuition was that connector hubs (brain regions that are diversely connected to multiple communities) would tend to be explained by pleiotropic genes, which encode the connectivity of multiple communities. To test this intuition, we examined whether each region's participation coefficient in the brain network was correlated with its participation coefficient in the gene encoding networks. The data support out hypothesis (**Fig. 3f,g**), suggesting that the genetic encoding of connector hubs mirrors their connectivity.

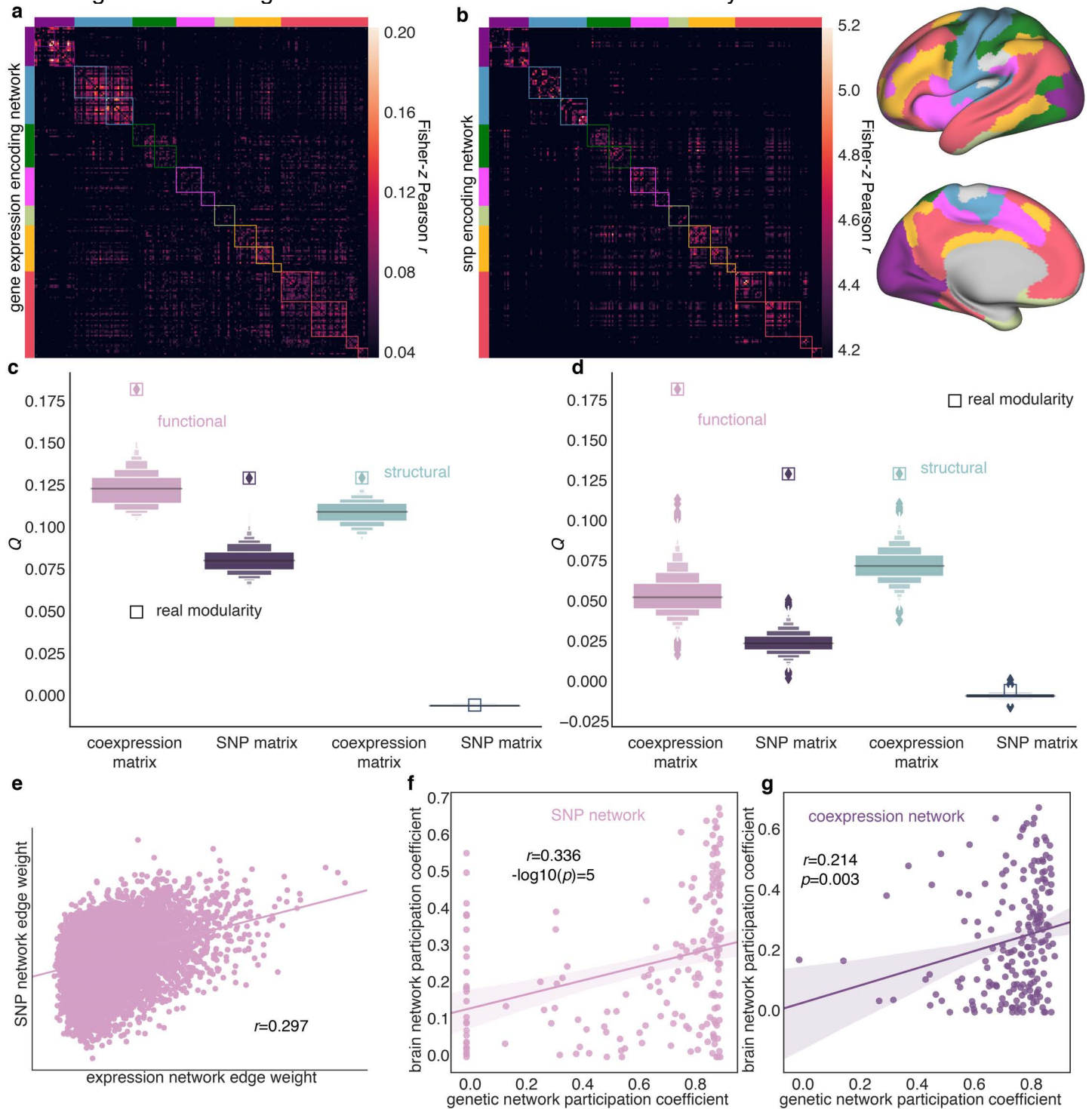

**Fig. 3 | The brain's network boundaries are genetically constructed via modular pleiotropy. a**, The gene coexpression encoding network. Each entry is the Pearson correlation coefficient $r$ between the number of times that we found each gene to fit each brain region's connectivity; if the connectivity of two regions is captured by the coexpression of the same genes, then they are connected strongly in this matrix. **b**, The SNP encoding network. Each entry is the Pearson correlation coefficient $r$ between the number of times that we found each gene's SNPs to fit the inter-subject variance in each region's connectivity; if the inter-subject variance in the connectivity of two regions is explained by the same gene's SNPs, then

they are strongly connected in this matrix. **c,d**, For two (**c** and **d**) spatially informed null models, the distribution of $Q$ values across null models, along with the real $Q$ value, is shown for functional and structural connectivity and a gene encoding network model at 15 percent density. Thus, both matrices in panels **a** and **b** are modular, even in comparison to the two spatially informed null model network partitions (one-sample t-test, across gene encoding network thresholds of 5,10, and 15 percent functional connectivity 582>t<1676, -log10(p)<100, dof=9999). **e**, The correlation between edge weights in the two networks; note that the correlation is $r=0.297$ (-log10(p)<5, dof=39599) for functional connectivity (shown) and $r=0.042$ for structural connectivity (not shown). **f,g**, In both the SNP encoding network and the gene coexpression encoding network, regions that function as connector hubs tend to also be connector hubs in the functional brain network. In each of these graphs, each data point is a brain region, and the region's participation coefficient in the genetic encoding network is plotted against its participation coefficient in the functional brain network (dof=191). Results were similar for structural networks ($r=0.24, 0.219$, $p=0.01, 0.002$, dof=191). These data suggest that the genetic signatures of connector hubs are diverse, in that they are similar to regions in multiple different communities.

To further test the hypothesis that brain connectivity is modularly encoded in the genome, we asked whether knowing a region's genetic signature—which genes capture a region's connectivity—either via gene coexpression or SNPs, is akin to knowing that region's role and community membership in the network. To address this question, we built a ridge regression learning model and used leave-one-node-out cross validation to predict each region's community membership as well as each region's nodal role as quantified by four graph metrics: strength, participation coefficient, within-community strength, and betweenness centrality (see Methods). In every prediction, we defined the features as the presence of which genes encode that region's connectivity. The predictions were especially strong for the participation coefficient **(Fig. 4a,b**, **Extended Data Fig. 11)**. SNPs did not predict a region's role in the structural brain network above chance. Moreover, predictions of a region's community membership were highly accurate for both structural and functional connectivity (**Fig. 4c,d**), but considerably more accurate for a region's functional network membership (69.4% for gene coexpression and 72.5% for SNPs), than for its structural network membership (55.4% and 14.2%, the latter of which is at chance level). Collectively, these results suggest that genes modularly capture network properties; different genes represent the connectivity of different types of nodes in the network.

Next, as a third way to test our modular hypothesis and to determine the contribution of individual genes to each community, we executed a PARIS (Probability Analysis by Ranked Information Score[41], see Methods). This analysis measures how strongly each gene encodes the connectivity of a single community via relative normalized mutual information. The RNMI values of many genes are statistically significant after FDR correction for multiple comparisons at a level of $q < 0.05$, demonstrating that particular genes capture the connectivity of single communities (**Extended Data Fig. 12**. and **Extended Data Fig. 13**). Notably, for both expression and SNPs, genes had higher RNMI values in the context of functional connectivity than in the context of structural connectivity, indicating stronger differential predictive power in function than structure (**Extended Data Fig. 12, Extended Data Fig. 13).** In sum, these results demonstrate that individual genes' SNPs and expression modularly represent functional network membership.

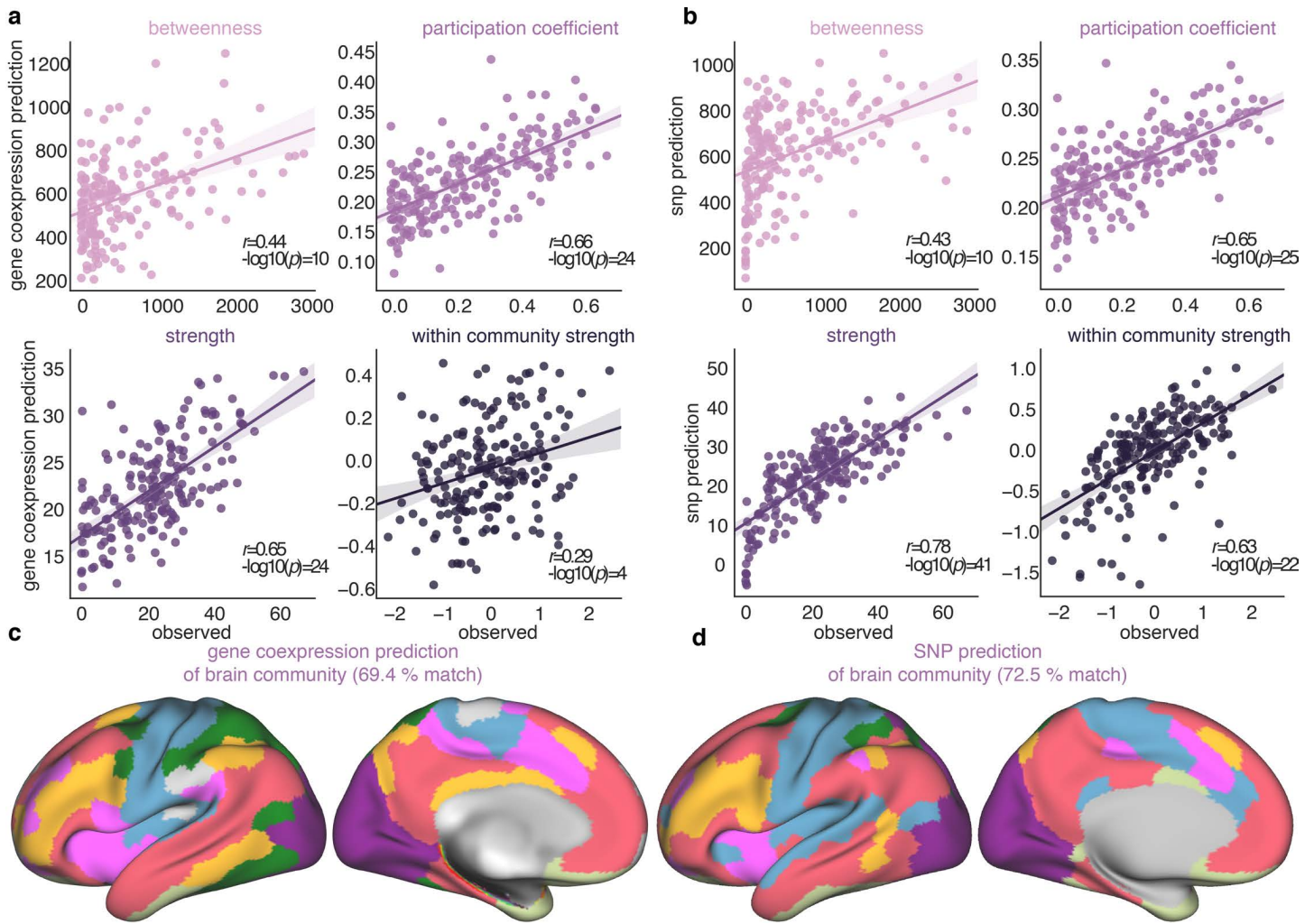

**Fig. 4 | Genetic signatures predict a region's role in the brain network and its membership in brain communities. a**, **b**, Prediction of four graph metrics based on which genes' coexpression (**a**) or SNPs (**b**) best explain a region's connectivity. **c**, **d**, Prediction of each node's community membership based on which genes' coexpression (**c**) or SNPs (**d**) best explain a region's connectivity. Note that chance for this prediction is 1/7 or 14.2 % (dof=191 for all panels).

## Linking gene coexpression, SNPs, brain connectivity, and behavior

We next turned to the question of whether the statistical link between brain connectivity and gene coexpression or SNPs is relevant for our understanding of human cognition and behavior. We began by assessing a region's *predictive value*, or the strength of the statistical relation between each node's connectivity and each behavior. Specifically, we calculated the mean correlation (across edges) between a given region's connectivity and a given behavior. Next, we calculate the correlation between a region's gene coexpression fit or a region's mean SNP fit across genes and its predictive value for each behavioral measure. For structural connectivity, we found no significant correlations between regional gene coexpression fits or SNP fits and regional predictive values for any behavioral measures after FDR correction for multiple comparisons (**Fig. 5a,b**). For functional connectivity, we found significant correlations between regional gene coexpression fits and SNP fits and regional predictive values for 10 and 6, respectively, of the 15 behavioral measures after FDR correction for multiple comparisons (**Fig. 5c**). These results suggest that, for certain behaviors, functional connectivity, but not structural connectivity, might mediate the relation between genetic variance and behavior, in that regions best fit by genetics are responsible for particular behaviors.

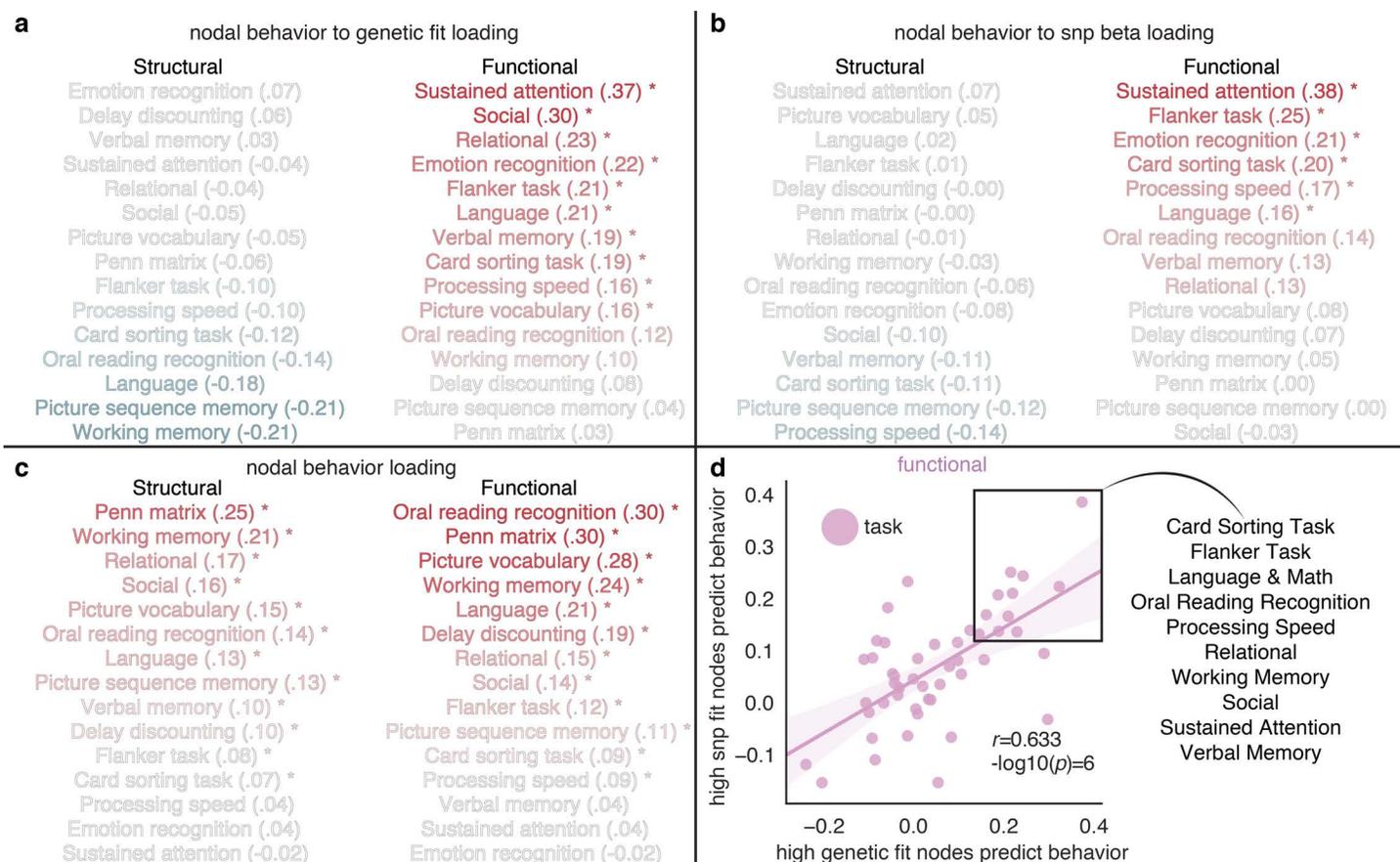

Fig. 5 | Specific behaviors depend on brain regions whose functional connectivity is well explained by gene expression and SNPs. a, The correlation between a node's gene coexpression fit and that same node's predictive value is shown for structural and functional connectivity. We mark significant correlations by asterisks ($p<0.05$ after FDR correction for multiple comparisons, dof=191). Correlations are only significant for functional connectivity. b, As in panel a, except that the calculation was executed with mean SNP betas for each node. c, For each behavioral measure, we fit a model using network properties of structural connectivity and functional connectivity, and a leave-one-out cross-validation procedure, to generate a prediction for the value of the behavioral measure for the left out subject (see Methods). Behavioral measures for which the model's predictions are significantly correlated with the true data are marked with an asterisk ($p<0.05$ after FDR correction, dof=191). d, On the y-axis, the degree to which a behavior depends on regions with connectivity that is well-fit by SNPs; on the x-axis, the degree to which a behavior depends on regions with connectivity that is well-fit by gene coexpression. Behaviors that depend on brain regions that are well fit by both gene coexpression and SNPs are listed (dof=48).

To ensure that the above results were not driven by a lack of a link between structural connectivity and behavior, we next built a predictive model by assigning network properties as features, and we used the model to predict the value of a behavioral measure for a given subject (see Methods). The models that were built from structural connectivity and the models that were built from functional connectivity significantly predicted many behavioral measures (**Fig. 5c**). Moreover, we observed that the accuracy of the models built from structural connectivity was highly correlated with the accuracy of the models built from functional connectivity ($r=0.771$, $p=0.003$, dof=48). Although the accuracy of the latter was greater than the accuracy of the former, the difference was not significant ($t=0.972$, $p=0.339$, dof=47). For these two analyses, to obtain a large enough $n$ for a student's $t$ or Pearson's $r$, we analyzed a larger set of 49 subject measures that include both the 15 behavioral measures and additional measures, like life satisfaction and personality factors, obtained via questionnaires.

Thus, while regions with high gene coexpression and SNP fits in the context of functional connectivity, but not structural connectivity, were predictive of behavior, both structural and functional connectivity are predictive of behavior. Accordingly, we hypothesized that distinct information about behavior is ensconced in function versus structure across brain regions. To test this hypothesis, we considered the correlation between the regions' predictive values estimated from structural connectivity and the regions' predictive values estimated from functional connectivity. We observed a weak and non-significant correlation between these two variables across the majority of behavioral measures ($p>0.05$, dof=191). To determine the specificity of this result, we also considered the correlation between the edges' predictive values estimated from structural connectivity and the edges' predictive values estimated from functional connectivity. We observed a weak and non-significant correlation between these two variables across all behavioral measures: the Pearson correlation coefficient $r$ was consistently less than or equal to 0.05. Thus, distinct information about behavior is ensconced in function versus structure across brain regions.

Finally, we discovered a link between gene coexpression and SNPs in the context of behavior and brain connectivity. For each behavior, we assess the degree to which that behavior loads onto regions with high gene coexpression or SNP fits. Specifically, we calculate the Pearson correlation coefficient $r$ between the regions' gene coexpression or SNP fits and the regions' predictive values for that behavior. We calculated this quantity separately for gene coexpression fits and SNP fits. Across tasks, we found a positive correlation between the amount that each behavior depends on regions whose functional connectivity is well fit by gene coexpression and the amount that each behavior depends on regions whose connectivity variance across subjects is well fit by SNPs (**Fig. 5d**). The behaviors that are best predicted by regions well fit by gene expression and SNPs demand higher-order cognition common to humans, including working memory, social cognition, and language abilities.

## Gene coexpression and SNPs differentially shape brain network architecture

Next, we sought to decipher the potential ways in which gene coexpression and SNPs differentially contribute to brain network architecture. We first sought to understand whether and how gene coexpression and SNP fits differentially relate to the heritability and conservation of brain connectivity, and to the evolutionary expansion of the cortex from macaques to humans. We used a previously generated map of how much regions have expanded from macaques to humans[34]. Given the conservation of brain gene expression across species[42], we reasoned that gene coexpression fits are likely to reflect constraints on brain architecture that are highly conserved across all individual animals of a given species and have not undergone evolutionary expansion. In contrast to such pervasive constraints, heritability reflects constraints (and perhaps innovations) that are specific to a single lineage of animals within a given species. Based on these intuitions, we expected to observe no significant relationship between the gene coexpression fits of connectivity and the heritability of connectivity, and we also expected that regions best fit by gene coexpression would not have expanded significantly from macaques to humans. Moreover, the connectivity of regions with high gene coexpression fits should be more conserved across subjects. In contrast, we did expect to observe a relationship between SNP fits and heritability, with the regions displaying the highest SNP fits being most heritable, because SNP fits track variation (and potentially innovation) across the population. Here, the connectivity of regions with high heritability and SNP fits should be less conserved across subjects.

To test these intuitions, we first calculated the statistical entropy (see Methods) in regional connectivity across subjects as a measure of conservation. Precisely in line with our predictions, we found that a region's gene coexpression fit is positively correlated with its conservation when both are estimated from functional connectivity, but not when both are estimated from structural connectivity, suggesting that gene coexpression fits of functional connectivity track conservation (**Fig. 6a,b**). Moreover, we found that a region's SNP fits are negatively correlated with its conservation when both are estimated from functional connectivity or structural connectivity, suggesting that SNP fits of functional and structural connectivity track inter-subject variation in regional connectivity (**Fig. 6a,b**).

We also found that the conservation estimated from functional connectivity is negatively correlated with the functional participation coefficient, but the conservation estimated from structural connectivity is positively correlated with the structural participation coefficient, suggesting that conservation differentially tracks functional and structural regional roles (**Fig. 6a,b**). Next, precisely in line with our prediction, we found that, in both structural and functional connectivity, gene expression fits were highest at regions that expanded the least during evolution (**Fig. 6a-c**). These evolutionarily expanded regions tended to be functional connector hubs and exhibit lower conservation in connectivity across subjects (**Fig. 6a-c**).

To determine the differential role of heritability in these relationships, we used the ACE twin model to estimate the heritability of each edge, controlling for age and sex. We found that functional connectivity was significantly more heritable than structural connectivity ($t = 31.113$, $-\log10$ ($p$)<5, dof=39599; **Fig. 6d**). Critically, while we observed that regional heritability of structural and functional connectivity were uncorrelated with regional gene coexpression fits, functional heritability was strongly positively correlated with regional SNP fits **Fig. 6a,b)** and regional functional participation coefficients (**Fig. 6a,b**), and were negatively correlated with regional functional conservation (**Fig. 6a,b**). Meanwhile, structural heritability was not correlated with regional structural participation coefficients (**Fig. 6a,b**) and was positively correlated with structural conservation (**Fig. 6a,b**). Finally, heritability was positively and negatively significantly correlated with cortical expansion in functional connectivity and structural connectivity, respectively, albeit weakly (**Fig. 6a,b**).

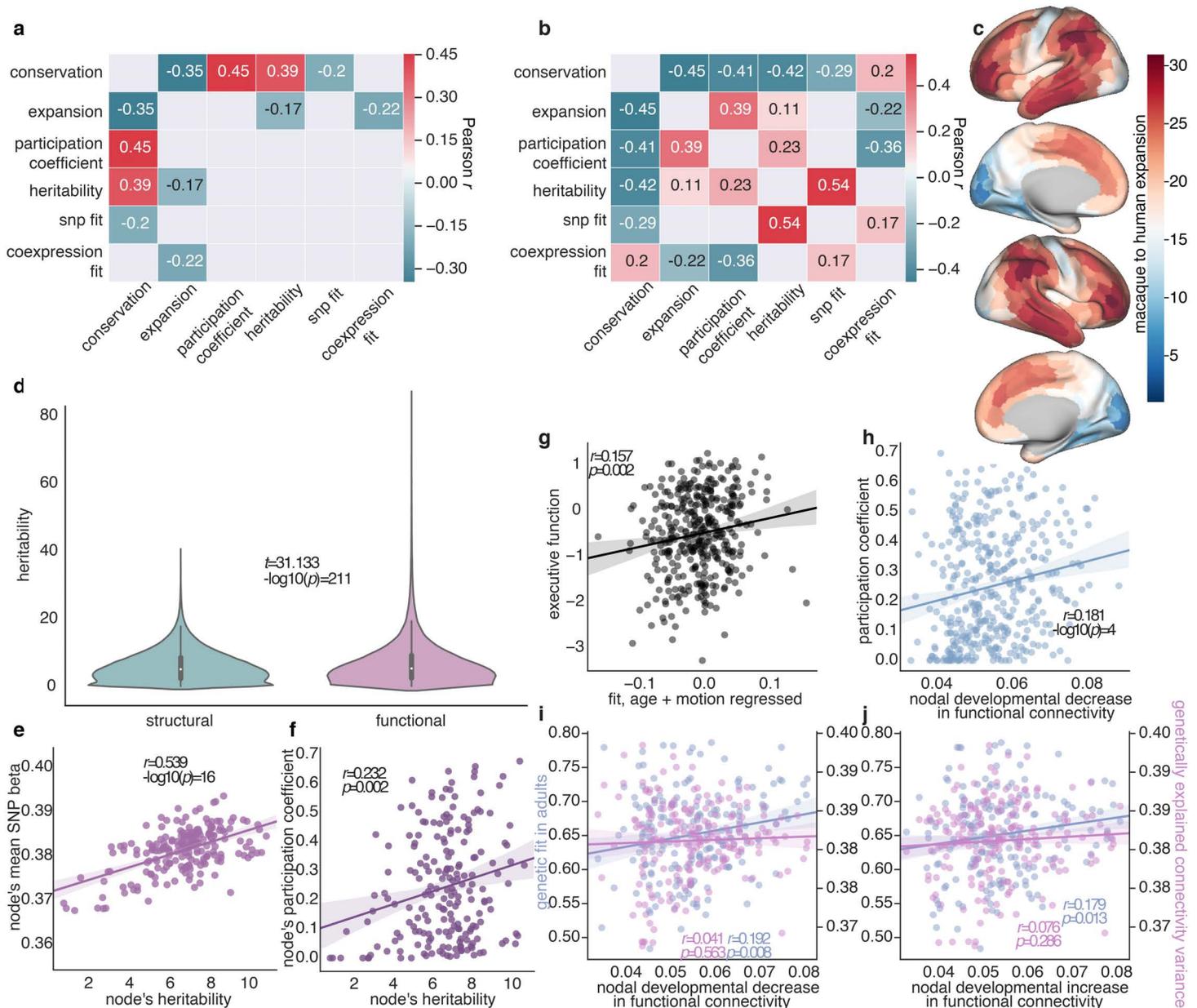

**Fig. 6 | Gene coexpression and SNPs differentially map to brain heritability, evolutionary expansion, conservation, connector hubs, and development**. **a,b**, For each region, we calculated the region's evolutionary expansion from macaque to human, the conservation of the region's connectivity across subjects, the heritability of the region's connections, the region's participation coefficient, and the region's gene coexpression and SNP fits. We then assessed the correlation between pairs of these measures. FDR multiple comparisons correction was applied at 0.05 and only significant correlations are shown; dof =191 for gene expression and SNP fit correlations, 398 for other correlations. **c**, Evolutionary expansion map from a previous analysis[34]; vertex values were mapped from right to left hemisphere in Freesurfer with xhemi, and the mean for each node was calculated. Red and blue areas are larger and smaller in humans compared to macaques, respectively. **d**, In the HCP data set, we calculated each functional and structural connectivity edge's heritability; functional connectivity was more heritable (dof=39599). **e,f**, We calculated each region's heritability of functional connectivity and then we calculated the correlation between those estimates and each region's SNP fit and functional participation coefficient (dof=191(e),398(f)). **g**, Using a large developmental sample, we show a scatterplot of the relation between executive function and the degree of similarity between individual functional connectivity matrices and the gene coexpression matrix after regressing out age and motion from the similarity values (dof=378). **h**, Scatterplot of the relation between a region's adult functional participation coefficient and the decrease in that region's functional connectivity strength throughout development (after regressing out motion) in the same large community sample (dof=398). **i,j**, Scatterplot of the relation between the gene coexpression fits or SNP fits of each region and regional increases (**i**) and decreases (**j**) in functional connectivity during development (after regressing out motion; dof=191).

## Gene coexpression and SNPs differentially shape brain network development

After examining gene coexpression and SNP fits in adult humans, we next turned to the question of whether and how those fits might inform our understanding of early adolescent brain development. We were particularly interested in identifying distinct developmental trajectories of regions with high and low gene coexpression fits for functional connectivity, and in determining their relevance for cognition. Thus, we began by generating a full region-by-region gene coexpression matrix by using each node's gene set and consequent gene coexpression row in the matrix. Then, we compared this gene coexpression matrix to the structural and functional connectivity matrices of adolescents between the ages of 8 and 13 years, as this is when the sharpest changes in connectivity[43]. We observed a positive correlation between a composite measure of executive function and the similarity between the functional connectivity matrix and the gene coexpression matrix, even after regressing out age and motion from the similarity measure ($r = 0.157$, $p = 0.002$, dof=378, **Fig. 6g**). In contrast, we observed a negative but insignificant ($r = -0.086$, $p = 0.092$, dof=378) correlation between the composite measure of executive function and the similarity between the structural connectivity and the gene coexpression matrix. Although our data are cross-sectional rather than longitudinal, these results support the notion that, the faster an adolescent's functional connectivity structure matches gene-coexpression, the faster the adolescent develops executive function.

Gene expression profiles of connector hubs could potentially reflect pruning in response to environmental experiences and heritable genetic innovations[23,44]. If true, one might hypothesize that connector hubs' edges (which recall have low gene coexpression fits) are potentially parsed away via pruning. In contrast, regional local nodes (which recall have high gene coexpression fits) might develop edges in a general genetically constrained manner. To test this hypothesis, we calculated the Pearson correlation coefficient *r* between age and each edge's weight, after regressing out mean framewise displacement from age. In line with our intuitions regarding pruning, regions with high functional participation coefficients tended to display a greater decrease ($r = 0.181$, $-\log_{10}(p)=4$, dof=398), but not increase ($r = -0.06$, $p = 0.229$, dof=398), in functional edge weights throughout development (**Fig. 6h**). Interestingly, regions with high functional gene coexpression fits tended to display both greater increases and decreases in edge weight than regions with low gene coexpression fits (**Fig. 6i,j**). No significant relationships between structural participation coefficients and structural edge weight changes were observed, and only increases in structural connectivity edge weight changes were significantly correlated with structural gene coexpression fits ($r = 0.171$, $p = 0.014$, dof=191). However, regions with strong *functional* gene coexpression fits tended to gain *structural* connections throughout development ($r = 0.211$, $p = 0.003$, dof=191). Interestingly, we did not observe any correlation between a region's SNP fit and its connectivity during development (**Fig. 6i,j**). Finally, we observed that regions most expanded during evolution prune, but do not gain, functional, but not structural, edges throughout development ($r=0.350$, $-\log_{10}(p)<5$, dof=398).

## Discussion

For brain and behavior, genomes have evolved through encoding brain phenotypes that execute optimal cognitive phenotypes: a brain with the cognitive disposition and ability to execute behaviors that typically increase fitness in the environment in which the animal is evolving. This mapping, from genes to brain network architecture to cognitive phenotypes, follows the principle of modular pleiotropy[28,29]. Previous work has uncovered a modular mapping between the brain's network architecture and behavior—each community executes a particular behavioral phenotype[4,11]—and we found that different sets of genes' expression and variance modularly map to each brain community. This discovery demonstrates that brain networks are boundaries in the brain that demarcate genetic transitions. Gene encoding is similar across cortical areas in the same community. However, gene encoding changes not as one move across the cortex but as one crosses a community border. However, there are specific locations, connector hubs, where these boundaries blur together. We discovered that pleiotropic genes from multiple mostly non-overlapping sets encode the connectivity of connector hubs. The genetic pleiotropy of these hubs mirrors their integrative role in brain function.

As connector hubs participate in cognitive processes that require multiple communities, it is not surprising that pleiotropic genes that encode different brain communities that underlie those cognitive process also encode the connectivity of connector hubs.

We also discovered a mapping between gene expression and single nucleotide polymorphisms. Single nucleotide polymorphisms at the genes whose coexpression mirrors the connectivity of a given brain region to the rest of the brain are associated with variation in that region's connectivity across subjects. Moreover, if the connectivity of two regions is mirrored by the coexpression of the same genes, then the regions' connectivity is encoded by the same genes' single nucleotide polymorphisms, and *vice versa*. This link also extends to behavior. Human-enriched cognitive tasks, including working memory, social cognition, and language abilities, similarly load onto regions with high gene coexpression and SNP fits.

We also found evidence consistent with the notion that functional connectivity is a stronger mediator between genotypes and cognitive phenotypes than structural connectivity. We found that functional connectivity is more strongly encoded in the genome than structural connectivity in terms of gene expression and SNPs. Moreover, functional, but not structural, nodes with high gene coexpression fits and SNP fits are the most predictive of cognitive and behavioral phenotypes. Critically, the information that functional connectivity provides about behavior is most strongly present at the regions that are most strongly encoded by gene expression and SNPs. Thus, when natural selection is selecting optimal phenotypes, functional connectivity may be one of the critical brain features that is shaped during this process. Under this hypothesis, natural selection is occurring on a genotype-phenotype-phenotype model in which certain genes (genotype) represent a brain region's connectivity, and variance at those genes leads to variance in brain connectivity (phenotype), as well as variance in cognition and behavior (phenotype). While the behavioral phenotype is selected, the brain phenotype, here, the brain's functional network architecture, is the form that is encoded in the genome. Critically, these discoveries between gene expression, SNPs, and behavior could aid in finding candidate genes for mental illness; if we know the regions disturbed in a given illness, then it is likely that single nucleotide polymorphisms at the genes' whose coexpression mirrors those regions' connectivity are involved in the illness.

We next differentially mapped gene coexpression and SNPs to heritability, functional conservation, participation coefficient, and evolutionary expansion of cortex. Regions that are well fit by SNPs are regions with functional connectivity that is highly heritable and lacks conservation across subjects. Moreover, the heritability of a region's functional connectivity, its lack of connectivity conservation across subjects, its participation coefficient, and its evolutionary expansion were all correlated with one other. In contrast, gene coexpression fits were unrelated to heritability, negatively correlated with the participation coefficient, and positively correlated with the conservation of a region's functional connectivity across subjects. Connector hub regions highly overlap with regions that are the most expanded during evolution[45]; while gene coexpression in the brain is mostly conserved across species[42], the human frontal cortex, where many connector hubs are located, shows marked differences in gene expression[46]. It is therefore interesting to speculate that local nodes with high conservation across subjects and high gene coexpression fits are highly constrained across *Hominoidae*, but that connector hubs with low conservation across subjects, low gene coexpression fits, and high heritability and SNP fits are the result of human specific innovations in connectivity. Consistent with these speculations, local nodes have non-human homologues, while many connector hubs are without clear non-human homologues[45]. Here, gene expression and SNPs might reflect different mechanisms underlying how functional connectivity mediates the relationship between genotypes and cognitive phenotypes. Finally, the results are also consistent with the notion that genetic control is more prominent for functional connectivity than for structural connectivity, as we observe that gene coexpression fits, SNP fits, and heritability are higher for function than for structure.

Finally, we differentially mapped gene coexpression and SNPs to brain development. We found that nodes with high gene coexpression fits showed stereotypical age-related changes in connectivity, slowly adding and pruning edges throughout development; in contrast, connector hubs, which are not well fit by gene coexpression, tend to prune edges throughout development. The results are consistent with the notion that connector hubs are likely pruning edges and optimizing the diverse connectivity that allows them to best integrate across the brain, control brain activity and connectivity, and support a modular functional architecture[4–6,47]. Moreover, we found that structural nodes with high gene coexpression fits gain edges during development and regions with high *functional* gene coexpression fits exhibit the greatest increases in *structural* connectivity throughout development. While no explicit direction can be inferred—structural changes might drive functional connectivity changes or *vice versa*—there is potentially a relationship between structure and function, where structural changes during development lead to functional changes[48]. While gene expression represents functional connectivity strongly and a plurality of structural connectivity configurations could give rise to this genetically encoded functional connectivity, structural connectivity underlies and constrains functional connectivity[49] and likely plays a role in how genetically encoded functional connections develop. Given that the diversity of connector hubs' connections actually increases throughout development[22], part of this process is likely genetically driven by a particular subset of genes' expression and SNPs. Finally, the faster the brain's functional connections develop to mirror adult gene coexpression, the faster executive function develops. In sum, gene coexpression is potentially encoding stable and conserved development trajectories (which nodes add or prune edges), while SNPs are encoding the actual final connectivity of each region, which can vary highly across subjects.

In summary, our results demonstrate the map between genes and the brain's network architecture exhibits principles of both modularity and integration, which is known as modular pleiotropy[28,29]. For each brain community, a mostly non-overlapping set of genes modularly encode the functional connectivity of that community. The integrative aspect of this encoding occurs at connector hubs, where pleiotropic genes from multiple mostly non-overlapping sets encode the connectivity of connector hubs. The functional communities here thus represent boundaries of genetic transitions and connector hubs are areas where these boundaries blur together. Functional connectivity is likely a stronger mediator between genes and cognition and behavior than structural connectivity. SNPs and gene coexpression likely differentially contribute to the brain's evolution, architecture, and development.

## Methods

**General methodological issues**. Several methodological considerations are pertinent to our work. Specifically, our analyses in general depend on diffusion imaging and probabilistic tractography accurately capturing white matter connectivity and on blood oxygen dependent changes (BOLD) fMRI accurately capturing statistical dependencies in population activity between groups of neurons. Ideally, both methods are identical in accuracy. While this ideal scenario is unlikely, the methods employed here to measure functional and structural connectivity are the state of the art and were applied to one of the most advanced and largest imaging datasets currently available. Moreover, these two methods are the only methods available to capture global human brain connectivity at millimeter resolution. It is also important to admit that while structural connectivity is a known physical property of the brain, the neural mechanisms underlying functional connectivity based on BOLD are not fully understood. Simply put, we are comparing a perfect measure of an under-defined phenomena (functional connectivity) to an imperfect measure of a very well defined phenomena (structural connectivity). Thus, a large amount of weight should not be placed on the reported differences between structure and function. Regardless, the functional connectivity analyses are of great interest alone and the structural connectivity results can be viewed as providing a rich comparison model. Finally, it is critical to note that our analyses sought to understand the macroscale principles of the mapping between gene expression and genetic variance and brain network architecture. However, more microscale analyses are indispensable. Future analyses could investigate how different categories or sets of SNPs, for example, based on molecular consequence or clinical significance, fit into the broad modular mapping between genetic variance and brain network architecture discovered here.

**Gene Expression Data.** Gene expression data were provided by the Allen Brain Institute. Across the six subjects, 1587 different anatomical locations were sampled using separate probes. To increase the stringency of our hypothesis testing, we focused on 16699 genes that had been previously identified as relevant for brain function[32]. We assigned each of the anatomical locations to one of 193 *a priori* defined parcels[37]. We only analyzed the left hemisphere, as only two of the six subjects had right hemisphere data. For each gene and probe of that gene, we subtracted the mean cortical gene expression (for that gene) of that probe from that probe[33]. We then calculated the mean cortical expression across probes for each gene. As multiple anatomical samples were contained within each parcel (i.e., brain region), we then calculated the mean expression of each gene across anatomical samples within each parcel. Collectively, these calculations generated a data matrix of size 193 (parcels) by 16699 (mean expression across probes in that parcel for a given gene). MNI space volumetric parcels were used here.

**SNP Data**. SNP data was provided by the Human Connectome Project. For each edge in the brain network thresholded at 5 percent (we used a thresholded matrix, because we wished to focus specifically on the variance of edges that are strong and present in most individuals), a linear GWAS analysis was executed in PLINK [39] using quality controls in line with a standardized protocol[40], including removing SNPs without a very high (95 %) genotype rate across individuals, SNPs with a low minor allele frequency (<0.01), and those deviating from Hardy-Weinberg equilibrium at -log10=5. The GWAS analyses were run separately for functional and structural connectivity. Each SNP was assigned to a single gene based on the hg19 gene-chromosome locations. Note that 13350 genes of the 16699 genes used for coexpression had SNP data.

**Adult human brain networks.** Functional and structural connectivity were estimated from images located in the Human Connectome Project S1200 release. All subjects with all four resting state scans and BedpostX diffusion data were processed. For both imaging modalities, we utilized a 400 parcel cortical parcellation that maximizes the similarity of functional connectivity within each parcel[37]. This parcellation also shows good agreement between parcellation boundaries and retinotopic boundaries, architectonic areas, especially for areas 3, 4, 2, and 17 in both hemispheres, and task activation maps[37], making it appropriate for analyzing both function and structure.

For functional connectivity, we analyzed the ICA-FIX resting state data, which removes noise signals via independent component analyses, as well as motion. Additionally, we removed the mean global signal and bandpass-filtered the time series from 0.009 to 0.08 Hz. Volumes with greater than 0.2 millimeters frame-wise displacement or a derivative root mean square (DVARS) above 75 were removed. Scans with less than 50 percent of frames left after motion artifact removal, were discarded. We used the MSMAll registration and calculated the mean activity time series of vertices on the cortical surface (fsL32K) in each parcel. Finally, we estimated the functional connectivity matrix for each subject as the pairwise Fisher z-transformed Pearson correlation coefficient $r$ between pairs of regional times series.

For structural connectivity, we used the BedpostX output calculated by the Human Connectome Project. The fsaverage5 space cortical parcellation was registered to the subject's native cortical white matter surface to serve as seeds and targets for probabilistic tractography run with the *probtrackx2* algorithm. After Markov Chain Monte Carlo sampling was used to build distributions on diffusion parameters at each voxel (BEDPOST), probabilistic tractography was executed (probtrackx2). For each parcel, we ran 1000 streams from each voxel in each parcel. The proportion of streams that reached each target was recorded and served as the measure of structural connectivity.

To assess the architecture within functional and structural connectivity matrices, we built network models of the data[50]. Specifically, we treated each brain region as a network node, and each structural or functional connection as a network edge. With this encoding, each connectivity matrix not only represents an adjacency matrix but also represents a brain network. By choosing a network model, we can access and utilize a wealth of computational tools and theories that have proven useful in recent work to understand the architecture of brain connectivity across spatiotemporal scales of measurement.

After calculating the mean matrices for structural connectivity and functional connectivity, we performed a series of additional calculations on the weighted networks represented by those connectivity matrices. The networks were thresholded at different densities to include between 1 and 10 percent of all possible connections in 50 equal steps. At each value of network density, we estimated the community structure using the InfoMap algorithm[51], which is based on describing the trajectory of a random walk on the network such that important structures have unique names, making our description of the random walk as succinct as possible. Intuitively, InfoMap is based on how information flows through the network; thus, the resulting communities are groups of nodes through which information flows the most frequently. In this way, the approach provides a useful alternative to maximizing the modularity index ($Q$) of the graph, as it takes into account the pattern of edges across nodes and how information could flow through the network. This process is then followed by calculations of the participation coefficient and within-community strength $z$-score[52,53]. Given a particular assignment of nodes to communities, the participation coefficient of each node can be calculated. Specifically, the participation coefficient *PC* of node *i* is defined as:

$$PC = 1 - \sum_{s=1}^{N_M} \left(\frac{K_{is}}{K_i}\right)^2,$$

where $K_i$ is the sum of node *i*'s edge weights, $K_{is}$ is the sum of node *i*'s edge weights to community *s*, and $N_M$ is the total number of communities. Thus, the participation coefficient is a measure of how evenly distributed a node's edges are across communities. A node's participation coefficient is maximal if it has an equal sum of edge weights to each community in the network. A node's participation coefficient is 0 if all of its edges connect it to a single community. The within-community-strength is calculated as:

$$WCS = \frac{K_i - \overline{CW_s}}{\sigma CW_s},$$

where $\overline{CW_s}$ is the mean edge weight of all nodes in community *s*, and $\sigma CW_s$ is the standard deviation of edge weights in community *s*. The variable $K_i$ is the edge weight between node *i* and all nodes in community *s*. We also calculated the strength and betweenness centrality of each node at every network density. Finally, we calculated the mean value for these four metrics—strength, betweenness centrality, participation coefficient, and within-community strength—across all network densities.

Much of the correspondence between gene coexpression and connectivity can be explained by the fact that a significant proportion of the variance in functional connectivity and gene coexpression can be explained by spatial proximity; the closer two regions are in the brain, the stronger their gene coexpression and the stronger their functional connectivity[35]. Thus, we regressed out Euclidian distance from structural and functional connectivity matrices for all analyses. Note, however, that graph metrics were extremely similar with and without distance regression: the participation coefficient was correlated at *r*=0.965 (-log10(*p*)<5, dof=398), the within community strength was correlated at *r*=0.928 (-log10(*p*)<5, dof=398), the betweenness was correlated at *r*=0.993 (-log10(*p*)<5, dof=398), and the strength was correlated at *r*=0.989 (-log10(*p*)<5, dof=398). Thus, while regressing out distance from the matrices prior to calculating graph metrics ensured that our results were not driven by distance, they do not prevent the intended and original interpretation of these graph metrics.

**Adolescence Brain Networks**. Functional connectivity matrices were calculated for subjects in the Philadelphia Neurodevelopmental Cohort. Given our analysis of executive function, we used the n-back task data. N-back functional images were processed using one of the top-performing pipelines for removal of motion-related artifact within the XCP engine[54,55]. Preprocessing steps included (a) correction for distortions induced by magnetic field inhomogeneities using FSL's FUGUE utility, (b) removal of the 4 initial volumes of each acquisition, (c) realignment of all volumes to a selected reference volume using MCFLIRT, (d) removal of and interpolation over intensity outliers in each voxel's time series using AFNI's 3DDESPIKE utility, (e) demeaning and removal of any linear or quadratic trends, and (f) co-registration of functional data to the high-resolution structural image using boundary-based registration to MNI space. The artifactual variance in the data was modeled using a total of 36 parameters, including the six frame-wise estimates of motion, the mean signal extracted from eroded white matter and cerebrospinal fluid compartments, the mean signal extracted from the entire brain, the derivatives of each of these nine parameters, and quadratic terms of each of the nine parameters and their derivatives. Both the BOLD-weighted time series and the artifactual model time series were temporally filtered using a first-order Butterworth filter with a passband between 0.01 and 0.08 Hz. Mean relative root-mean-square framewise displacement was calculated for each subject for nuisance regression against connectivity changes during development. As in the human connectome project, we estimated the functional connectivity matrix for each subject as the pairwise Fisher z-transformed Pearson correlation coefficient *r* between pairs of regional times series. For structural connectivity, we used the same BEDPOST and probtrackx2 methods described above to construct structural connectivity matrices.

**Machine Learning Algorithm**. We developed a novel algorithm to learn the set of *n* genes that maximized the gene coexpression fit (*g*) of each node *a*:

$$g_a = pearsonR(x, y),$$

where *x* is node *a*'s gene expression similarity with all other nodes and *y* is node *a*'s structural or functional connectivity to all other nodes. The gene coexpression fit was calculated separately for structural and functional connectivity. The simulated annealing process began with *n* (15-200) random genes. The starting temperature was set to 0.5, and the system was cooled in increments of 0.001 degrees. The temperature set the fraction of genes in the set of *n* genes to remove before recalculating the gene coexpression fit. Thus, at a temperature of 0.5, one half of the genes in the *n* gene set were removed and replaced with other genes not in the set. The genes were chosen with *numpy's* random.choice. At each temperature, the appropriate number of genes was chosen for replacement, and *g* was recalculated. At each temperature, 100 iterations were run, and the best *g* and the corresponding *n* genes were chosen and kept if the value of *g* was higher than that observed at the previous temperature. This process concluded at a temperature of 0.01. All of these parameters were set *a priori* and were not changed to increase or decrease performance.

For all values of *n*, coexpression was not significantly higher when using the Pearson correlation coefficient *r* than when using the Spearman correlation coefficient. This consistency in findings demonstrates that the former estimate is not unduly driven by outliers (**Extended Data Fig. 1**). Moreover, for all values of *n*, we confirmed that the gene coexpression fit values calculated with Spearman correlation coefficients and Pearson correlation coefficients *r* were highly correlated (**Extended Data Fig. 2**). Thus, for subsequent analysis, we chose to use the Pearson's correlation coefficient because it requires no ordinal transform of the data and because gene expression is interval. Additionally, we found that the gene coexpression fit values obtained for any single value of *n* were highly correlated with gene coexpression fit values obtained for any other value of *n*. Significant correlations ($0.40 \geq r \leq 0.82$, $p <$ -log10($p$) = 3, dof=191) were found between all *n* genes analyzed, with *n*=50 exhibiting the highest minimum *r* value (**Extended Data Fig. 3**). Thus, using 50 genes generates gene coexpression fits that are most representative of the gene coexpression fits across all values of *n* genes. For the developmental analysis, we use an *n*=50 genes, given that it was the most representative *n* gene set and this analysis required using only a single *n* genes. However, for all analyses, we used the mean gene coexpression fit across *n* genes or counted the number of times a gene fit a region across *n* genes.

We calculated the number of regions for which a given gene's expression fits the region's structural or functional connectivity. The distribution of these values was heavy tailed for functional connectivity, but not for structural connectivity, with some genes fitting as many at 22 regions in functional connectivity (**Extended Data Fig. 7**). However, as our results demonstrate, these 22 regions tend to be in the same community.

Finally, we executed a gene ontology enrichment analysis with GOrilla[36]. For both structural and functional connectivity, separately, we ranked genes by the number of times each gene was selected to fit brain connectivity. The genes most commonly found to fit brain connectivity are preferentially involved in known cellular component, biological process, and molecular function mechanisms. Critically, for both structural and functional connectivity, many gene ontologies relevant to brain function and structure were found to be significantly enriched after correcting for multiple comparisons (**Extended Data Fig. 8, Extended Data Fig. 9**).

**Gene-Nodal Role Predictive Model.** In every prediction, we defined the features to be the number of times (across *n* genes) that a gene was found to display coexpression that fit the region's connectivity or genes' SNP betas for that region's connectivity variance. We refer to the former as a gene coexpression prediction, and we refer to the latter as a SNP prediction. Features were generated for functional connectivity and for structural connectivity separately. Given the large number of features (~17,000) we used Tikhonov regularization to prevent overfitting and a ridge-regression learning model in scikit-learn. A leave-one-node-out cross validation was used, and we compared the predicted with the real values for each node's network role or community membership.

**PARIS.** As a third way to test our modular-pleiotropy hypothesis and to determine the contribution of individual genes to each community, we executed a PARIS (Probability Analysis by Ranked Information Score[41]) analysis, which ranks the similarity of gene profiles to a target profile using a mutual information metric. More specifically, we define a target vector that is a binary vector of size 1-by-n where n is the number of brain regions under study, which intuitively indicates regional participation in a given network community. We also define a gene profile vector that is a weighted vector of size 1-by-n, indicating that this gene was used in predicting a region's connectivity. With this target vector and gene profile vector, we calculate the *rescaled normalized mutual information* between the target vector and each gene profile vector. Recall that the normalized mutual information is defined as $NMI(t,x) = \frac{MI(t,x)}{H(t,x)}$, and intuitively reflects the mutual information normalized by the joint entropy. This value is then rescaled by the target's own NMI and attached a sign based on the direction of correlation to finally give the rescaled normalized mutual information:

$$RNMI(t,x) = sign(\rho(t,x)) \frac{NMI(t,x)}{NMI(t,t)},$$

where *x* is the gene profile, *t* is the target vector, and $\rho(t,x)$ is the correlation coefficient. Code for the software is freely available as a GenePattern module (www.genepattern.org) or on the GenePattern Module Archive (http://www.gparc.org/).

For gene coexpression, each gene was assigned the value indicating how many times that gene fit the connectivity of each region across *n* runs. For SNPs, each gene was assigned the absolute beta coefficient indicating how well SNPs at that gene were found to fit variance in connectivity for each region. For each gene, the PARIS calculation measures relative normalized mutual information (RNMI), which is intuitively the extent to which the above values differentiate between a single community and other communities. For example, if the absolute beta coefficients of particular genes' SNPs are high for regions within a single community and low for regions in all other communities, the RNMI will be near 1 for those genes. Beta values that are more uniformly distributed across regions in all communities will have an RNMI near 0. Thus, we can interpret the RNMI as reflecting the gene's differential predictive power for a community. For each community, we show the top 15 genes for structural and functional connectivity separately in **Extended Data Fig. 12a,b**. and **Extended Data Fig. 13a,b**. Critically, the RNMI values of many of the genes are statistically significant after FDR correction for multiple comparisons at a level of $q < 0.05$, demonstrating that particular genes capture the connectivity of single communities. Interestingly, for both gene coexpression and SNPs, the visual, motor, and default communities (communities that are not dominated by connector hubs) contained more genes with high RNMI values than the other communities, suggesting strong differential predictive power in functional connectivity. Notably, for both expression and SNPs, genes had higher RNMI values in the context of functional connectivity than in the context of structural connectivity, indicating stronger differential predictive power in function than structure (**Extended Data Fig. 12c,d, Extended Data Fig. 13c,d).** In sum, the PARIS results demonstrate that individual genes' SNPs and expression modularly represent functional network membership.

Finally, even though we observe less heterogeneity in the genes whose coexpression represents each region's functional connectivity relative to structural connectivity (**Extended Data Fig. 7**), we also found that genes represent functional connectivity in a more modular manner than they represent structural connectivity. This apparent contradiction is explicitly resolved by the PARIS analysis. In the context of functional connectivity, the Spearman correlation coefficient between the community index (0-6) for which the mean RNMI is the highest for a gene and the community index (0-6) for which the gene most frequently fits connectivity is $r$=0.73 (p<-log10(p)=30, dof=191). Thus, even though a gene can be chosen for multiple regions when fitting gene coexpression to functional connectivity, these multiple regions are more likely to be in the same community than in different communities.

**Gene Coexpression Null Models** We developed two spatially informed random network null models in order to validate the gene coexpression fit results: a random gene selection null model and a shuffled gene expression null model. In the random gene selection null model, we randomly selected genes for which to calculate gene coexpression. In the shuffled gene expression null model, the gene expression values are shuffled across genes and nodes, and then random selection (identical to the first model) is applied. For both models, functional or structural connectivity of each node is compared to the gene coexpression matrix.

All real fits are significantly higher than fits obtained in the random gene selection null model for structure (59 < $t$ > 112, -log10($p$)<5, dof=192) and for function (41 < $t$ > 75, -log10($p$)<5, dof=192) (**Extended Data Fig. 4**). All real fits are significantly higher than fits obtained in the shuffled gene expression null model for structure (91 < $t$ > 197, -log10($p$)<5, dof=192) and for function (99 < $t$ > 165, -log10($p$)<5, dof=192) (**Extended Data Fig. 4**). These data demonstrate that our algorithm actually increases the fit between coexpression and connectivity.

In the shuffled gene expression null model, we did not observe any difference between functional and structural connectivity fits across the number of genes (-0.640 > $t$ < 2.1, $p$ >0.05, dof=192) (**Extended Data Fig. 4**), demonstrating that there is nothing inherent about structural connectivity that drives the fits to be lower than functional connectivity. However, in the random gene selection null model, the fits for functional connectivity were consistently higher across the number of genes (3.071 < t > 3.10, $p$= 0.003, dof=192) (**Extended Data Fig. 4**). These data indicate that regardless of the genes that are selected for gene coexpression, gene coexpression fits functional connectivity better than it fits structural connectivity.

In the random gene selection model, we found a negative correlation between fits and participation coefficient for both structural and functional connectivity (**Extended Data Fig. 5**); in other words, gene expression in general does not appear to fit participation coefficient well. However, in the shuffled gene expression null model, there was no significant correlation between fits and the participation coefficient (**Extended Data Fig. 6**). These data suggest that there is nothing about the distribution of gene expression that makes fitting connector nodes more difficult.

These null models ensured that nothing about the distribution of gene expression or connectivity strengths makes it more difficult to fit gene coexpression to structural connectivity than to functional connectivity, or to fit gene coexpression to one type of node (e.g., connector hubs) than to another type of node.

**Spatial Null Gene Encoding $Q$ Models**. Two spatial nulls models were utilized to test $Q$ values of the gene coexpression and SNP encoding networks. First, with a random initialization of 17 random nodes into 17 different communities, we iteratively assign each node (in random order) to the community of the node it is closest to in physical space. Second, given that the array of canonical community indices is ordered in a spatially contiguous manner (e.g., all visual center community nodes are in the array from position 0 to 12, visual periphery community nodes are in position 13 to 24, and so on), rolling (explicitly, np.roll) the array a random amount moves the boundaries across the cortex, but maintains communities that contain nodes that are spatially contiguous. For each null model, 10000 iterations and corresponding $Q$ values were generated.

**Predictive Model of Behavior** We built a support vector machine (**LinearSVR** in *scikit-learn*) supervised learning model to predict subjects' task performance. Our model's features ($n=4$) assessed how well node's participation coefficients, within-community strengths, and network connectivity (as operationalized by the edge weights) and modularity (Newman's modularity quality index $Q$[56]) are optimized for task performance. Model features were calculated in an identical manner. For example, for the feature that measures how well the subjects' participation coefficients are optimized for performance, for each brain region, we calculated the Pearson correlation coefficient $r$ across subjects between each region's participation coefficient and task performance. The feature, then, across subjects, is the Pearson correlation coefficient $r$ between the subject's regions' participation coefficients and the former Pearson $r$-values that represent how well that region's participation coefficients correlate with performance. The same procedure is executed for within-community strengths and edge weights. Finally, we used the $Q$ values of the network. Note that these features are identical to a previous model of network connectivity, except that that model used features from task data and resting state data and a simpler support vector machine is used in place of deep neural networks[6]. Using a leave-one-out cross-validation procedure, we built the model features and then fit the model on all subjects except one. The model was then used to predict the left-out subject's task performance.

**Entropy** Entropy was calculated with: scipy.stats.entropy. We interpret this entropy value as a measure of conservation, as it reflects the average amount of information conveyed across events, when considering all possible outcomes. If connectivity is highly conserved across subjects, a lot of information is conveyed by each subject's connectivity, and entropy is high.

**Letter Value Boxenplots** Distributions were plotted use **seaborn.boxenplot** with stock parameters. Letter-value plots are a variation of boxplots that replace the whiskers with a variable number of letter values, selected based on the uncertainty associated with each estimate and hence on the number of observations. Any values outside the most extreme letter value are displayed individually. These two modifications reduce the number of "outliers" displayed for large data sets, and make letter-value plots useful over a much wider range of data sizes. Letter-value plots remain true to the spirit of boxplots by displaying only actual observations from the sample, and remaining free of tuning parameters. Letter-value plots are based on the letter values, with one box for each pair of lower and upper letter values. The median is shown by a vertical line segment, and the innermost box is drawn at the lower and upper fourths, as in the conventional boxplot. An incrementally narrower box is drawn at the lower and upper eighths, and a narrower one still at the lower and upper sixteenths. Boxes with matching heights correspond to the same depths[57].

**Data Availability Statement:** All data analyzed here are publicly available via the Human Connectome Project, The Philadelphia Neurodevelopmental Cohort, and the Allen Brain Institute.

**Code Availability Statement:** All analysis code can be found at: https://github.com/mb3152/modular_pleiotropy

## Acknowledgments

We thank Maja Bućan and Jing Zhang for their assistance in executing the GWAS analyses in PLINK. MAB acknowledges support from 5 T32 MH 106442-3, Ruth L. Kirschstein National Research Service Award. DRR acknowledges support from NIH R01MH102609. TDS AND DSB acknowledge support from NIH R01MH113550. DSB acknowledges additional support from the John D. and Catherine T. MacArthur Foundation, the Alfred P. Sloan Foundation, the ISI Foundation, the Paul Allen Foundation, the Army Research Laboratory (W911NF-10-2-0022), the Army Research Office (Bassett-W911NF-14-1-0679, Grafton-W911NF-16-1-0474DCIST- W911NF-17-2-0181), the Office of Naval Research, the National Institute of Mental Health (2-R01-DC-009209-11, R01-MH112847, R01-MH107235, R21-M MH-106799), the National Institute of Child Health and Human Development (1R01HD086888-01), National Institute of Neurological Disorders and Stroke (R01 NS099348), and the National Science Foundation (BCS-1441502, BCS-1430087, NSF PHY-1554488 and BCS-1631550). The content is solely the responsibility of the authors and does not necessarily represent the official views of any of the funding agencies.

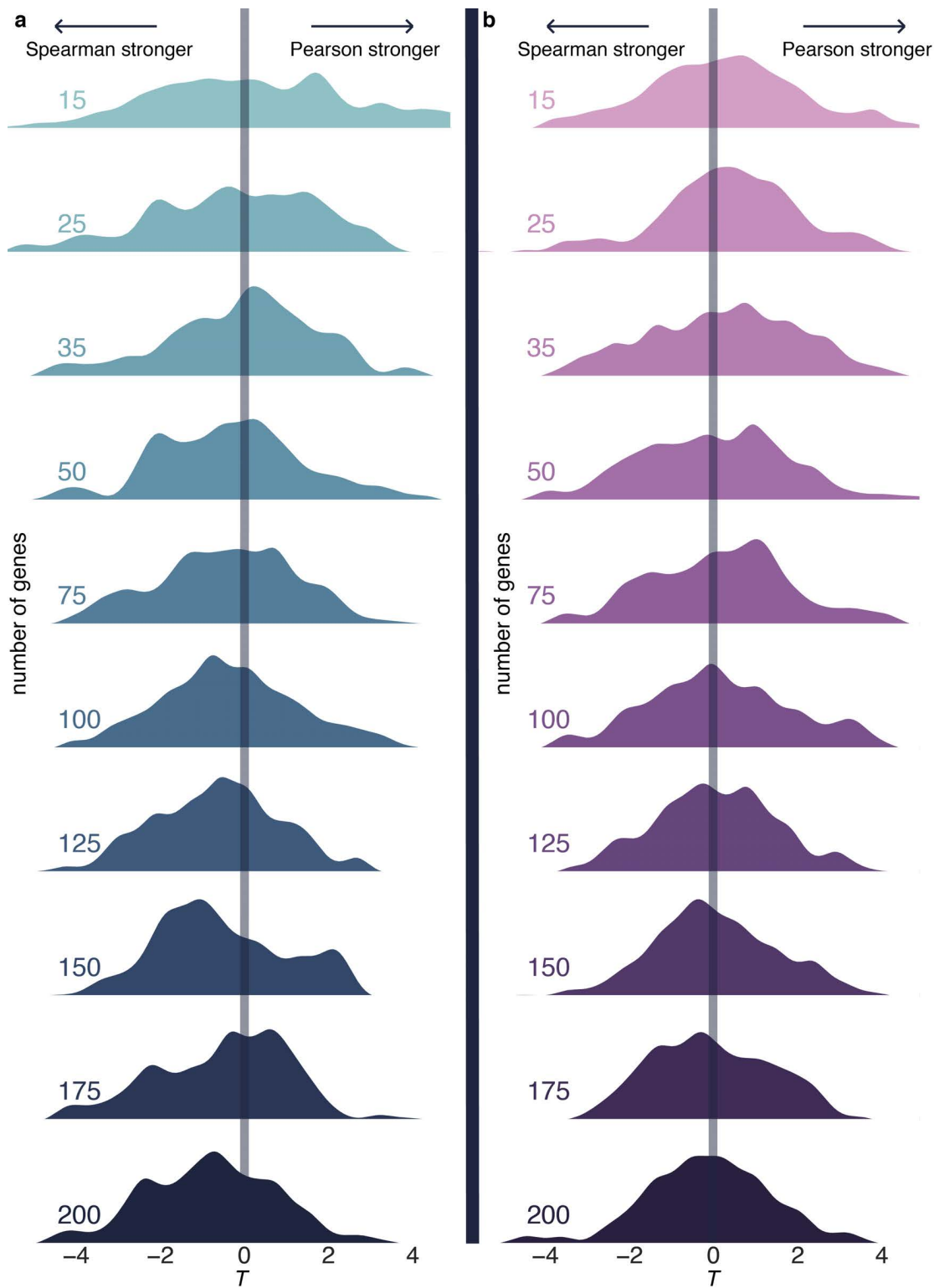

**Extended Data Fig. 1 | Measuring gene coexpression with the Pearson correlation coefficient is robust to outliers**.
**a**, Gene coexpression was fit to structural connectivity using 15 to 200 genes. For each number of genes and for each node's coexpression values, a student's *t*-test (dof=192) was applied to determine any differences between the gene-coexpression values calculated with the Pearson's *r*-value *versus* the Spearman's *rho*-value. Higher gene-coexpression values for Pearson's *r* could potentially reflect the fact that outliers in the two expression arrays are driving a high gene coexpression value. For all numbers of genes, coexpression was not significantly higher when using Pearson's *r* than when using Spearman's *rho*, demonstrating that measuring gene coexpression with Pearson's *r* is not unduly driven by outliers in this data. The distribution, across nodes, of those *t*-values is shown for each *n* genes. The mean *t*-value was *t* = 0.105. **b**, As in panel (**a**), but for functional connectivity. The mean *t*-value was *t* = -0.353.

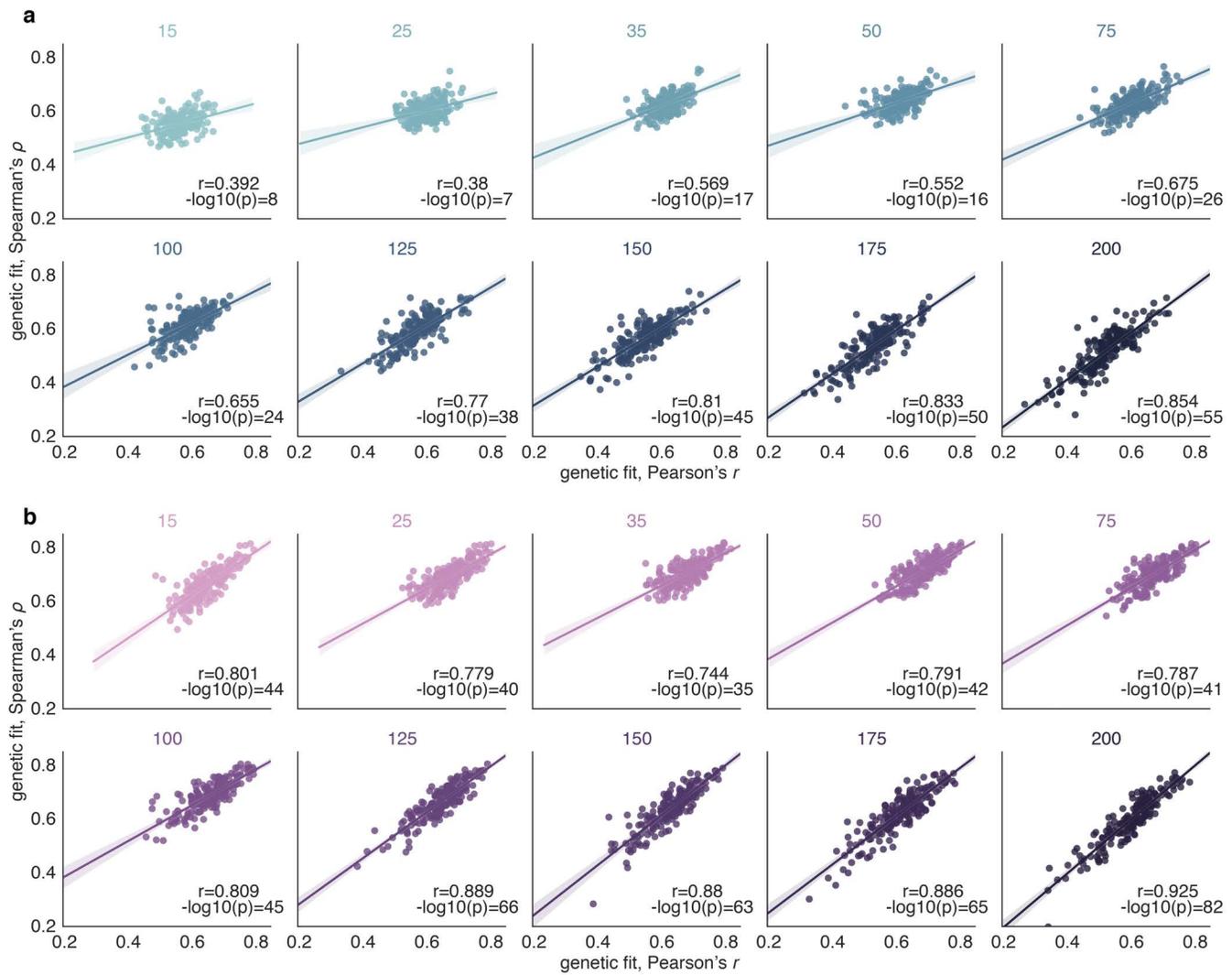

**Extended Data Fig. 2 | Gene coexpression fits to connectivity are similar when gene coexpression is calculated with a Pearson's *r* or a Spearman's rho**. **a**, For each number of genes, the Pearson's correlation coefficient between structural connectivity gene coexpression fits when the gene coexpression was calculated using either the Pearson's *r*-value (x-axis) or the Spearman *rho-value* (y-axis). **b**, As in panel (**a**), but for functional connectivity gene coexpression fits. The dof=191 for all panels.

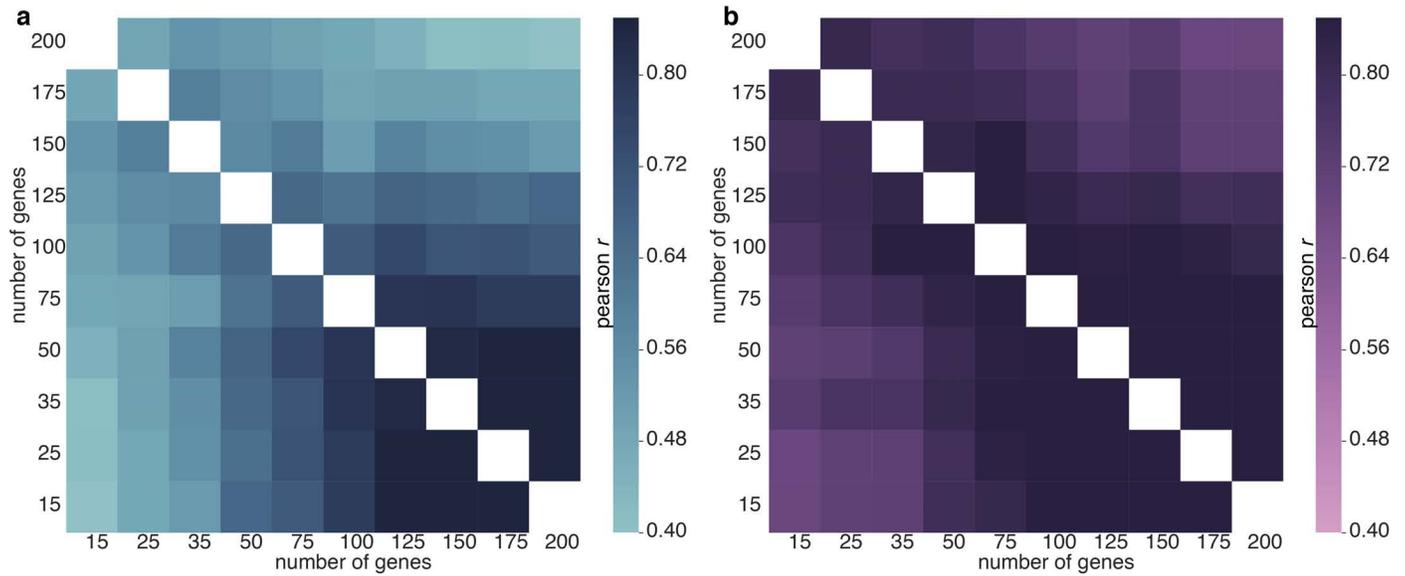

**Extended Data Fig. 3 | The set of 50 genes that are most representative of gene coexpression fits of sets with other numbers of genes. a,b,** The correlation of gene coexpression fits across regions between sets of genes with different numbers of genes (15-200) for structural (**a**) and functional (**b**) connectivity. The minimum (across rows or columns) of correlations between a set of genes and other sets of genes was highest for 50 genes for both structural and functional connectivity. The dof = 191 for all panels.

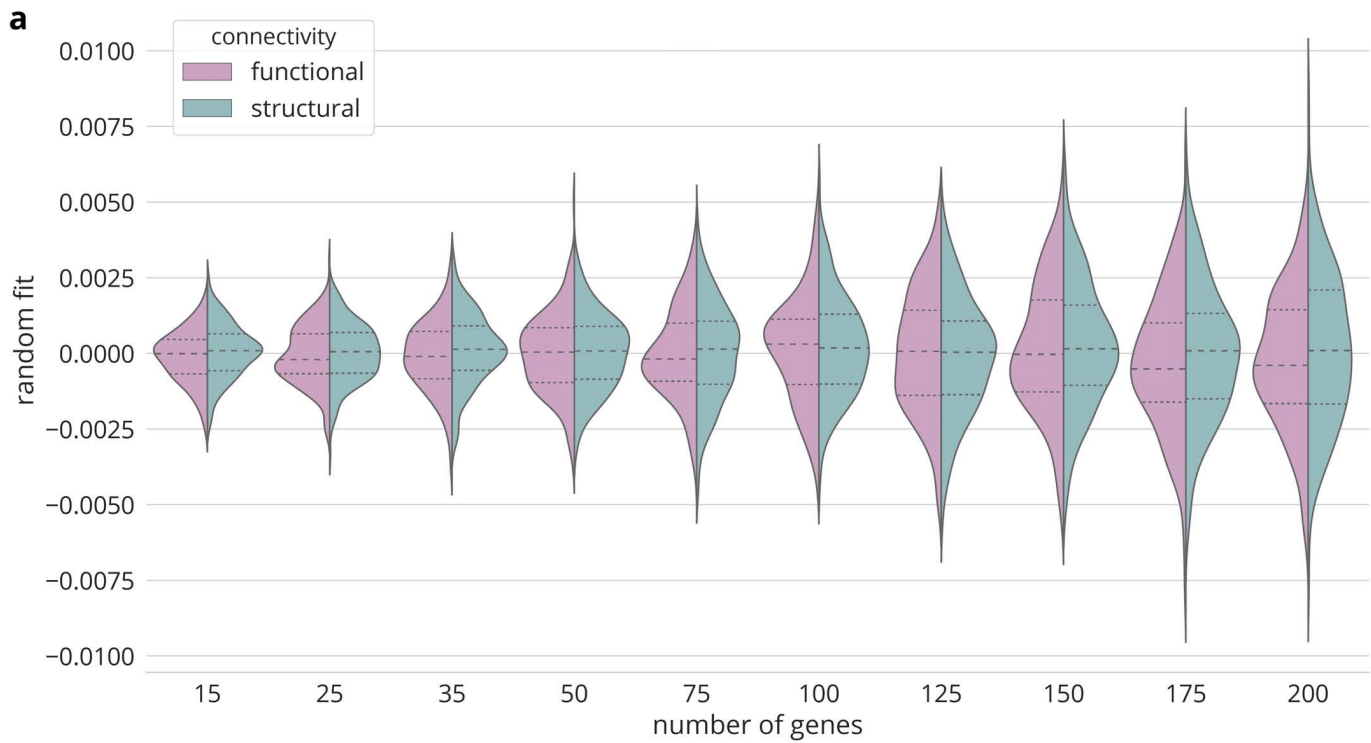

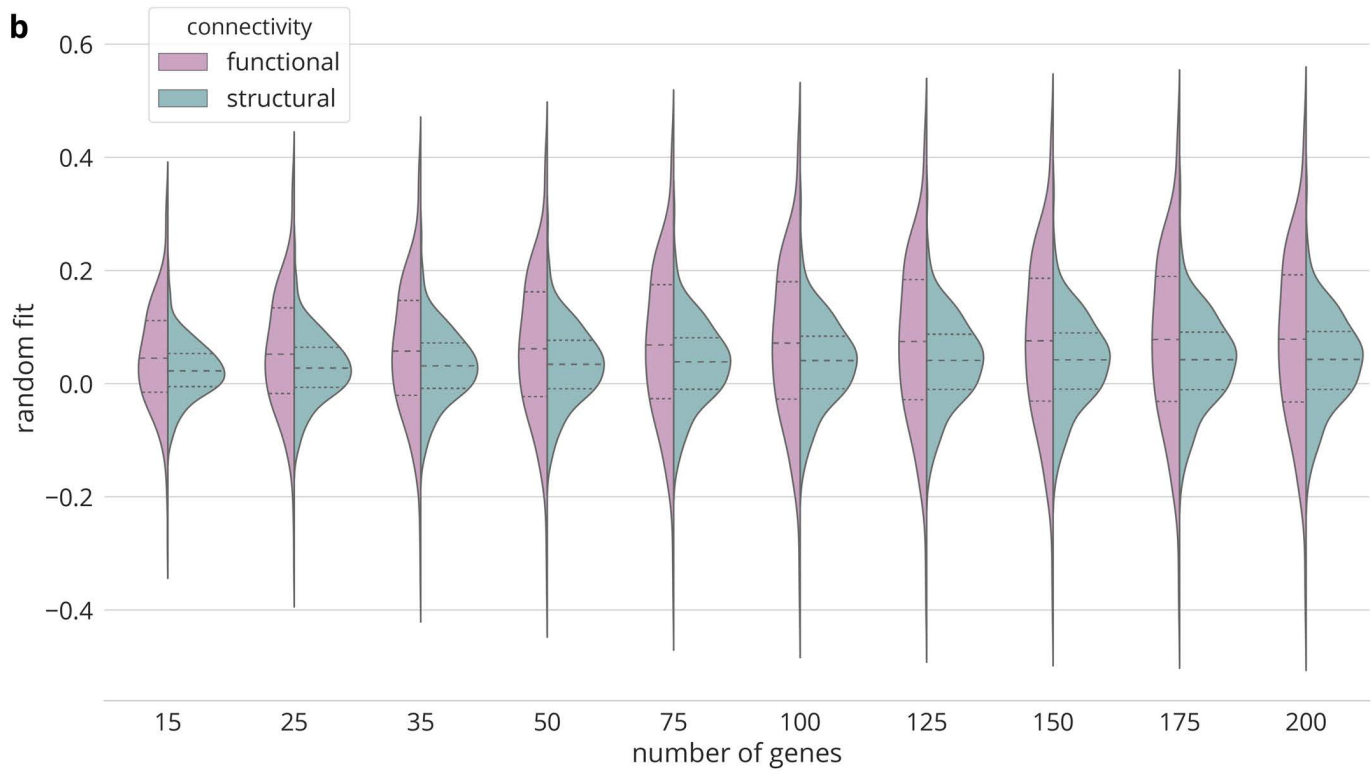

**Extended Data Fig. 4 | Two spatially informed random network null models of gene coexpression fits**. **a**, Gene coexpression fits in the shuffled gene expression null model, where the gene expression values are shuffled across genes and regions, and then random selection is applied. **b**, Gene coexpression fits in the random gene selection null model, where we randomly selected genes for which to calculate gene coexpression. All real fits are significantly higher than the fits obtained in the shuffled gene expression (**a**) null model for structure (91< $t$ >197, -log10($p$)<5) and for function (99 < $t$ > 165, -log10($p$)<5) and for the random gene selection null model (**b**) for structure (59 < $t$ >112, -log10($p$)<5) and for function (41 < $t$ > 75, -log10($p$)<5). In the shuffled gene expression null model (**a**), we did not observe any differences between functional and structural connectivity fits across the number of genes (-0.640 > $t$ < 2.1, $p$ > 0.05). However, in the

random gene selection null model (**b**), the fits for functional connectivity were consistently higher across the number of genes (3.071 < t > 3.10, $p$ < 0.003). The dof = 192 for all panels.

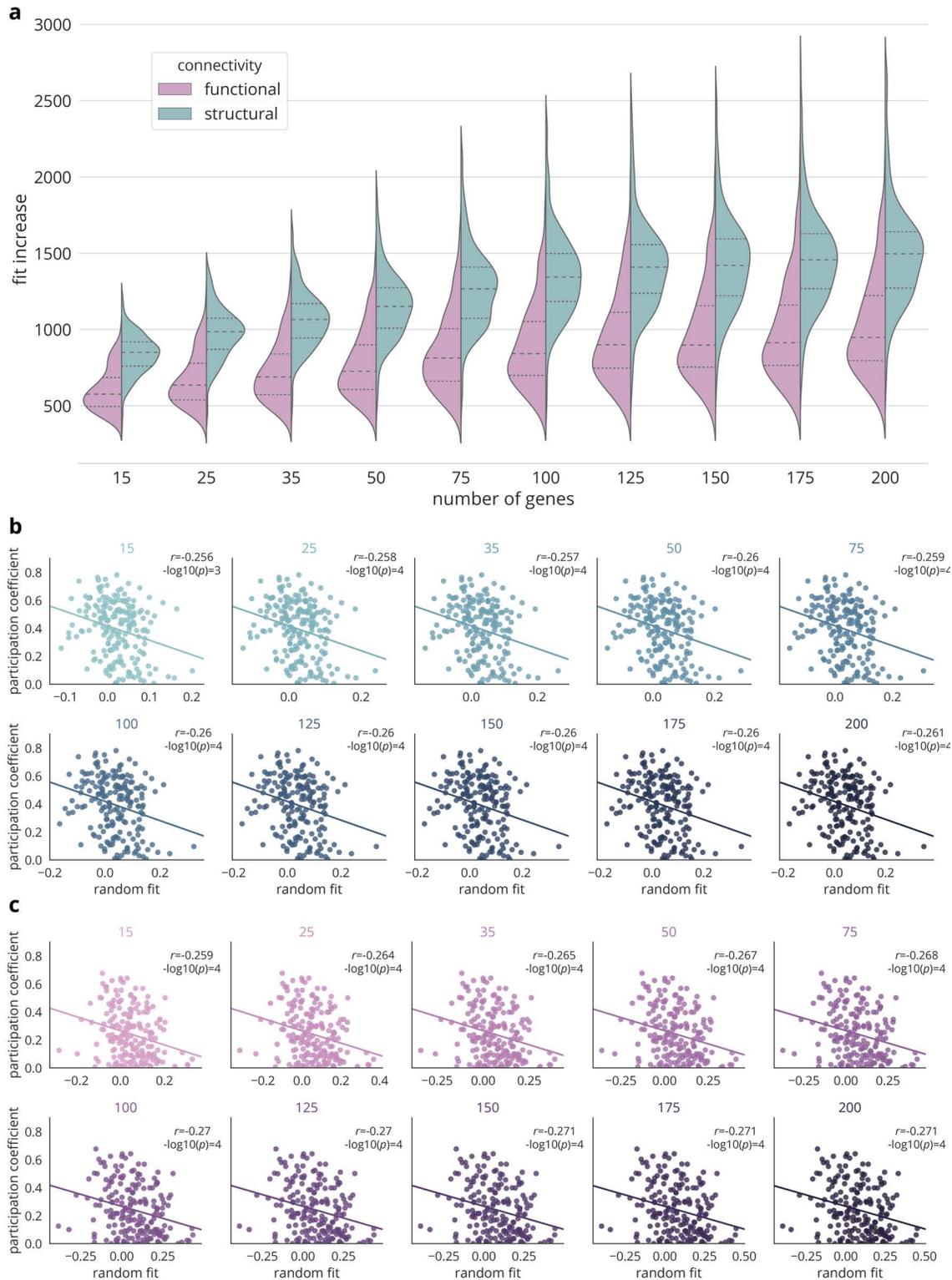

**Extended Data Fig. 5 | Gene coexpression fits are higher using simulated annealing to find genes than using a random gene null model**. **a**, We measured the degree to which the mean gene coexpression fit in the real model differed from that obtained in the random gene null model by performing a one sample $t$-test (here, fit increase) between the distribution of means in the null models and the mean in the real model. Across all $n$ genes and nodes, we found that by selecting particular genes for coexpression via simulated annealing, we were able to increase the gene coexpression fits of structural connectivity to a greater extent than functional connectivity (dof=9999). **b,c**, The correlation between the

participation coefficient and each region's mean gene coexpression fit in the random gene null model in structure (**b**) and function (**c**); the dof = 191.

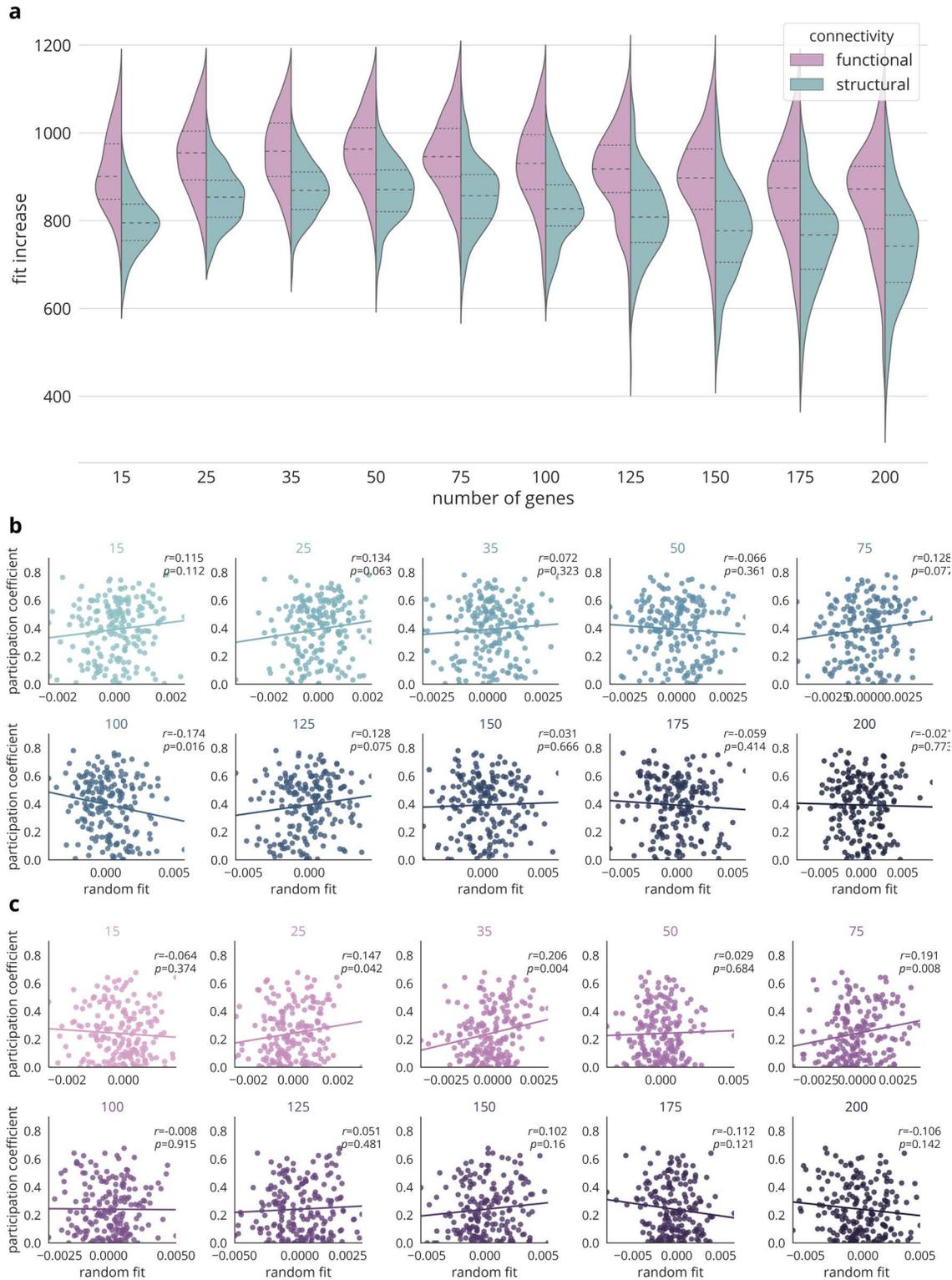

**Extended Data Fig. 6 | Gene coexpression fits are higher using simulated annealing to find genes than using a shuffled gene null model**. **a**, We measured the degree to which the mean gene coexpression fit in the real model differed from that obtained in the shuffled gene null model by performing a one sample *t*-test (here, fit increase) between the distribution of means in the null models and the mean in the real model. Across all *n* genes and nodes, we found that by selecting particular genes for coexpression via simulated annealing, we were able to increase the gene coexpression fits of functional connectivity to a greater extent than structural connectivity (dof=9999). **b,c**, The correlation between the participation coefficient and each region's mean gene coexpression fit in the shuffled gene null model in structure (**b**) and function(**c**); the dof = 191.

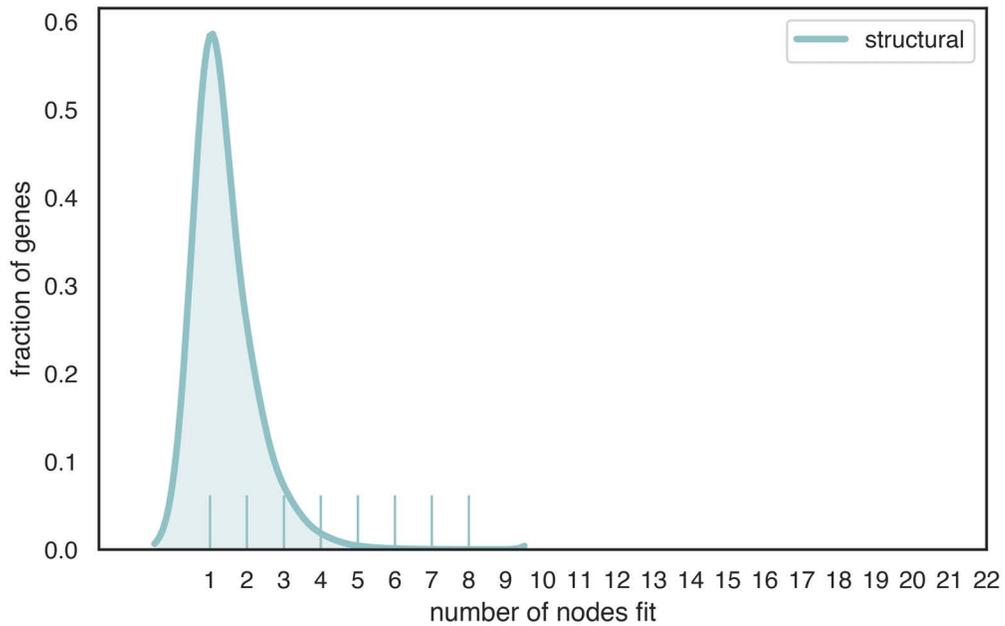

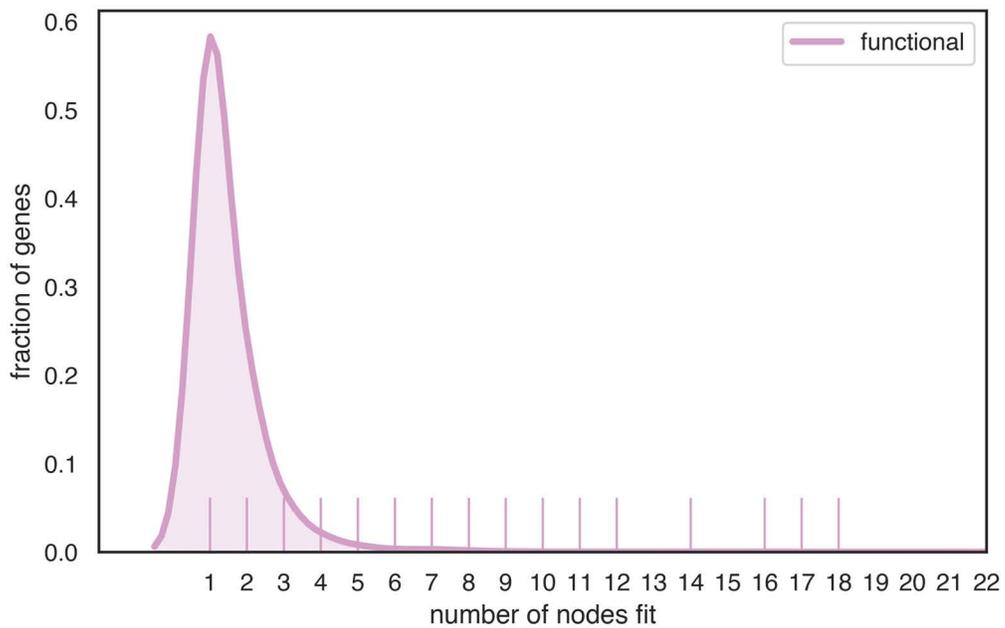

**Extended Data Fig. 7 | Distribution of the fraction of genes that are found to fit the connectivity of multiple brain regions.** For each region, 50 genes are found to fit the region's structural or functional connectivity. Across regions, some genes are found more regularly. Genes are more regularly selected for multiple regions in functional connectivity than in structural connectivity, evidenced by the heavier tail of the distribution, for which the Kolmogorov-Smirnov test for goodness of fit compared to a normal distribution is 22.21 for functional connectivity, and 7.15 for structural connectivity.

| Process | Component | Function |
|---|---|---|
| Defense response (1.53) ** | Extracellular space (1.58) ** | Signaling receptor activity (1.71) ** |
| Inflammatory response (2.43) ** | Extracellular region (1.76) ** | Molecular transducer activity (1.67) ** |
| G-protein coupled receptor signaling pathway (1.73) ** | Extracellular matrix (2.83) ** | Transmembrane signaling receptor activity (1.75) ** |
| Regulation of signaling receptor activity (2.02) ** | Extracellular region part (1.55) ** | Receptor regulator activity (2.2) ** |
| Cell adhesion (2.06) ** | Collagen-containing extracellular matrix (2.6) * | Receptor ligand activity (1.88) ** |
| Biological adhesion (2.05) ** | | G-protein coupled receptor activity (1.58) ** |
| Extracellular matrix organization (2.47) ** | | G-protein coupled receptor binding (2.21) * |
| Multicellular organismal process (1.55) ** | | Scavenger receptor activity (3.08) * |
| Extracellular structure organization (2.13) ** | | Peptidase regulator activity (1.92) * |
| Positive regulation of leukocyte migration (2.9) ** | | Peptidase inhibitor activity (2.37) * |
| Cellular developmental process (1.27) ** | | Cytokine activity (2.54) * |
| Granulocyte migration (2.86) ** | | Hormone activity (3.65) * |
| Regulation of response to external stimulus (2.06) ** | | |
| Developmental process (1.42) ** | | |
| Regulation of cell adhesion (2.14) ** | | |
| Cellular response to interleukin-1 (2.13) ** | | |
| Detection of stimulus (1.56) ** | | |
| Detection of chemical stimulus (1.64) * | | |
| Regulation of immune effector process (2.22) * | | |
| Granulocyte chemotaxis (2.73) * | | |
| Response to interleukin-1 (1.9) * | | |
| Positive regulation of cytokine production (1.42) * | | |
| Myeloid leukocyte migration (2.42) * | | |
| Immune response (1.34) * | | |
| Involved in sensory perception detection of chemical stimulus (1.62) * | | |
| Peptide secretion (3.16) * | | |
| Coupled to cyclic nucleotide second messenger g-protein coupled receptor signaling pathway, (2.38) * | | |
| Signaling (1.53) * | | |
| Neutrophil chemotaxis (2.72) * | | |
| Neutrophil migration (2.61) * | | |
| Positive regulation of interleukin-6 production (2.04) * | | |
| Anatomical structure development (1.49) * | | |
| Regulation of interleukin-6 production (1.82) * | | |
| Cell-cell signaling (1.51) * | | |
| Cell differentiation (1.3) * | | |
| Taxis (2.25) * | | |
| Regulation of muscle system process (7.9) * | | |
| Chemotaxis (2.25) * | | |
| Involved in sensory perception detection of stimulus (1.55) * | | |
| Coupled receptor signaling pathway adenylate cyclase-modulating g-protein (2.59) * | | |

**Extended Data Fig. 8 | GOrilla analysis of genes whose coexpression fits functional connectivity.** Genes were ranked according to the number of times they were discovered by our algorithm to fit gene coexpression to functional connectivity. All gene ontologies for which the FDR corrected *p*-values were lower than 0.05 are shown and marked with an asterisk. Gene ontologies for which the FDR corrected *p*-values were lower than 0.01 are marked with two asterisks. Gradations in color represent FDR corrected *p*-values, with darker text having lower *p*-values.

| Process | Component | Function |
|---|---|---|
| Involved in sensory perception detection of chemical stimulus (1.52) ** | Extracellular region (1.41) ** | G-protein coupled receptor activity (1.46) ** |
| Positive regulation of immune system process (1.71) ** | Extracellular space (1.66) ** | Signaling receptor activity (1.47) ** |
| Detection of chemical stimulus (1.51) ** | Secretory granule (2.64) ** | Molecular transducer activity (1.44) ** |
| Positive regulation of immune response (1.86) ** | Extracellular region part (1.47) ** | Transmembrane signaling receptor activity (1.31) ** |
| Cell differentiation (1.34) ** | Secretory vesicle (2.2) ** | Receptor regulator activity (1.81) ** |
| Immune response (1.64) ** | Extracellular matrix (3.12) ** | Receptor ligand activity (2.02) ** |
| Positive regulation of immune effector process (2.09) ** | Plasma membrane (1.22) * | Olfactory receptor activity (1.56) ** |
| Regulation of immune effector process (2.92) ** | Collagen-containing extracellular matrix (3.28) * | Growth factor receptor binding (2.99) ** |
| Defense response (2.03) ** | Endoplasmic reticulum lumen (2.23) * | Oxygen binding (3.16) * |
| Regulation of calcium ion binding (644.9) ** | Integral component of plasma membrane (1.7) * | Signaling receptor binding (1.71) * |
| Negative regulation of calcium ion binding (644.9) ** | | Oxygen carrier activity (8.91) * |
| In sensory perception of smell detection of chemical stimulus involved (1.56) ** | | G-protein coupled peptide receptor activity (1.7) * |
| Detection of stimulus (1.38) ** | | Interleukin-1 receptor activity (12.25) * |
| Regulation of signaling receptor activity (1.85) ** | | |
| Involved in sensory perception detection of stimulus (1.45) ** | | |
| Regulation of immune response (2.4) ** | | |
| Regulation of cell activation (2.05) ** | | |
| Cell-cell adhesion (1.9) ** | | |
| Regulation of immune system process (1.58) * | | |
| G-protein coupled receptor signaling pathway (1.25) * | | |
| Regulation of lymphocyte chemotaxis (6.21) * | | |
| Developmental process (1.22) * | | |
| Extracellular matrix organization (7.34) * | | |
| Cellular developmental process (1.22) * | | |
| Regulation of leukocyte activation (2.03) * | | |
| Response to molecule of bacterial origin (2.1) * | | |
| Regulation of defense response (2.15) * | | |
| Inflammatory response (2.47) * | | |
| Multicellular organismal process (1.22) * | | |
| Sequestered calcium ion into cytosol regulation of release of (7.38) * | | |
| Extracellular structure organization (3.67) * | | |
| Positive regulation of leukocyte migration (3.45) * | | |
| Gas transport (8.0) * | | |
| Regulation of sequestering of calcium ion (5.61) * | | |
| Regulation of peptidyl-tyrosine phosphorylation (2.76) * | | |
| Immune system process (1.43) * | | |
| Positive regulation of cytokine production (1.62) * | | |
| Plasma-membrane adhesion molecules cell-cell adhesion via (2.16) * | | |
| Response to external biotic stimulus (1.34) * | | |
| Response to lipopolysaccharide (2.07) * | | |
| Regulation of lymphocyte differentiation (2.33) * | | |
| Regulation of leukocyte differentiation (2.31) * | | |
| Regulation of lymphocyte activation (1.5) * | | |
| Oxygen transport (8.91) * | | |
| Response to biotic stimulus (1.33) * | | |
| Regulation of leukocyte chemotaxis (2.75) * | | |
| Regulation of response to external stimulus (1.54) * | | |
| Membrane raft organization (6.27) * | | |
| Cellular response to biotic stimulus (2.24) * | | |
| Negative regulation of calcium ion import (214.97) * | | |
| Regulation of chemotaxis (2.09) * | | |
| Regulation of endothelial cell apoptotic process (4.13) * | | |
| Anatomical structure development (1.25) * | | |
| Regulation of response to stimulus (1.37) * | | |
| Cell adhesion (1.51) * | | |
| Negative regulation of lymphocyte activation (1.88) * | | |
| Calcium ion homeostasis (3.89) * | | |
| Humoral immune response (2.18) * | | |

**Extended Data Fig. 9 | GOrilla analysis of genes whose coexpression fits structural connectivity**. Genes were ranked according to the number of times that they were discovered by our algorithm to fit gene coexpression to structural connectivity. All gene ontologies for which the FDR corrected *p*-values were lower than 0.05 are shown and marked with an asterisk. Gene ontologies for which the FDR corrected *p*-values were lower than 0.01 are marked with two asterisks. Gradations in color represent FDR corrected *p*-values, with darker text having lower *p*-values.

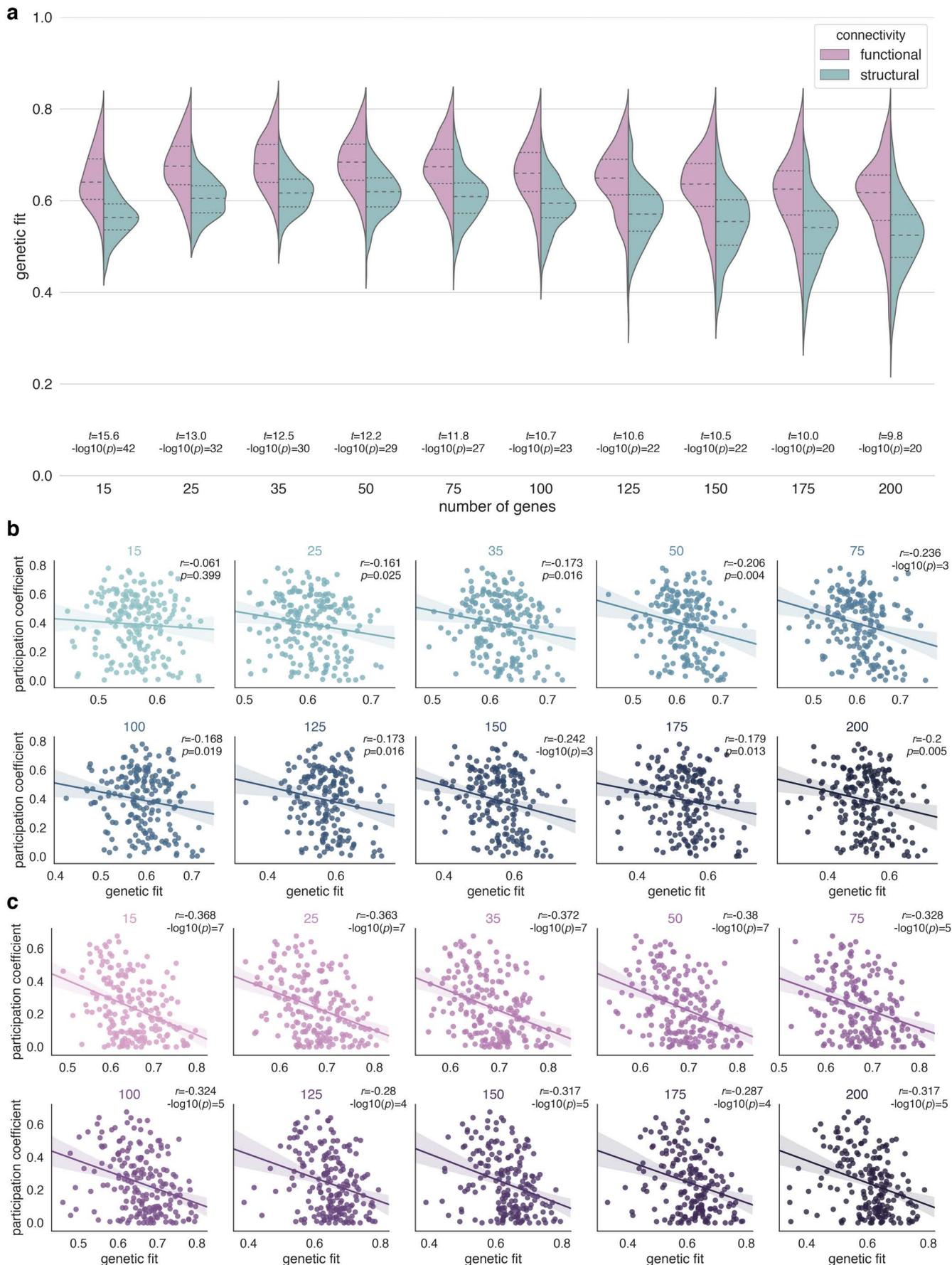

**Extended Data Fig. 10 | Gene coexpression fits are stronger for functional connectivity than for structural connectivity and correlate with the participation coefficient. a**, Gene coexpression fits for functional and structural connectivity across all sets of *n* genes analyzed, where *n* is the number of genes used to calculate gene coexpression.

For all *n* genes, gene coexpression fits were stronger for functional connectivity than for structural connectivity (dof=192). **b**, The correlation between structural gene coexpression fits and structural participation coefficients for all *n* genes. Besides *n* genes = 15, the correlations were significant for all *n* genes (dof=191). **c**, The correlation between functional gene coexpression fits and functional participation coefficients for all *n* genes (dof=191). Correlations were significant for all *n* genes.

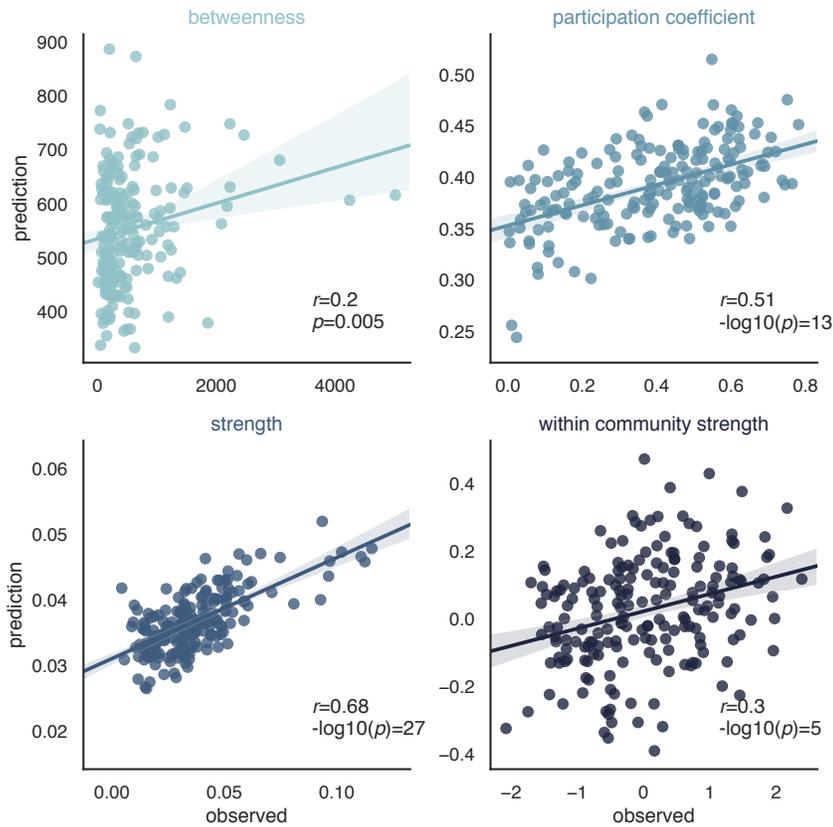

**Extended Data Fig. 11 | Predictions of structural connectivity graph metrics using gene coexpression.** As in Figure 4 for functional connectivity, for four structural connectivity graph metrics—betweenness, participation coefficient, strength, and the within community strength—we predicted each graph metric based on which genes' coexpression best explains a region's structural connectivity. The dof = 191 for all panels.

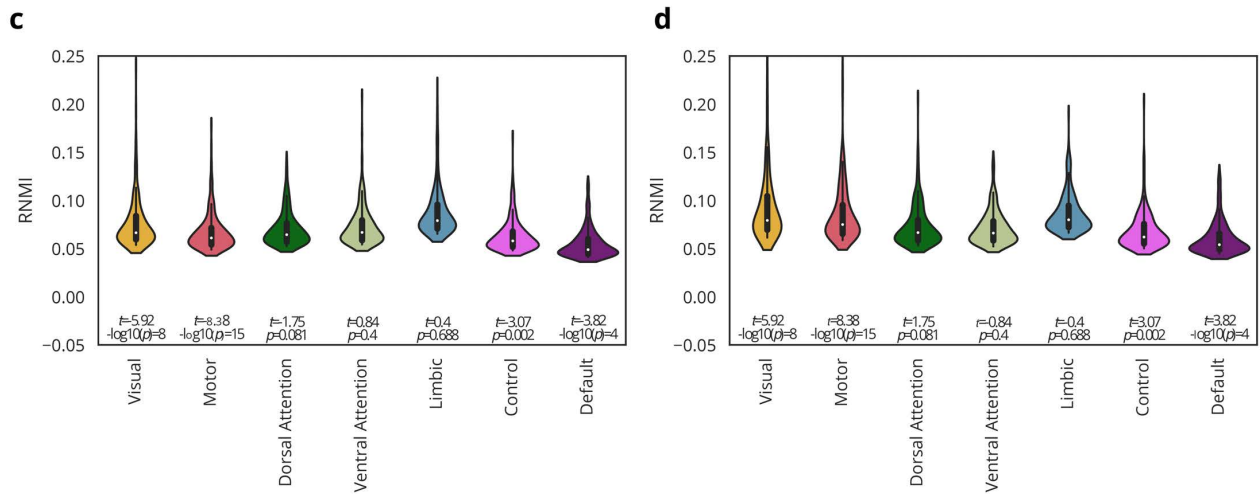

**Extended Data Fig. 12 | PARIS analysis of genes whose expression represents brain connectivity. a,b,** PARIS was used to find genes that are differentially useful in predicting connectivity for a particular community versus all other communities. The top 15 genes for (**a**) structural and (**b**) functional connectivity are shown with the FDR corrected *p*-value for each community. **c,d,** Violin plots of the top 250 genes RNMI values for functional (**c**) and structural (**d**) connectivity. Note that, in both functional and structural connectivity, visual and motor communities have the most genes that strongly differentiate them from the other communities. Also note that in functional connectivity, the motor, visual, control, and default mode communities have genes that strongly differentiate communities to a greater extent than structural connectivity does (all communities, $t = 8.11$, $-\log 10(p) = 19$, dof = 192, individual community comparisons in panels (**c**) and (**d**)), suggesting that gene coexpression represents functional connectivity in a more modular fashion than it represents structural connectivity.

## a

| Visual | Motor | Dorsal Attention | Ventral Attention | Limbic | Control | Default |
|---|---|---|---|---|---|---|
| TNNT2, 0.239* | RADIL, 0.179 | PSMA4, 0.144 | CUST_4_PI417507815, 0.208 | FAP, 0.219 | CLCN6, 0.166 | TRAPPC2L, 0.120 |
| HIST1H3J, 0.194* | WIF1, 0.155 | NOTCH4, 0.137 | FAM153B, 0.168 | HBS1L, 0.200 | HLA-DRB5, 0.124 | TNFAIP3, 0.116 |
| C13orf39, 0.168 | CCDC105, 0.138 | AC093283.3, 0.132 | TMEM160, 0.143 | AFAP1-AS1, 0.188 | TREM2, 0.119 | A_24_P577289, 0.112 |
| GPR26, 0.160 | OAZ1, 0.135 | SPANXN4, 0.127 | RFX1, 0.142 | HP1BP3, 0.181 | C10orf137, 0.113 | TSPAN2, 0.112 |
| FSCN3, 0.154 | SLC16A14, 0.123 | A_24_P919706, 0.125 | A_24_P944472, 0.126 | SERPING1, 0.170 | LIMCH1, 0.112 | LINC00238, 0.107 |
| RGS4, 0.151 | C20orf74, 0.123 | OR51G2, 0.118 | TCP11L2, 0.121 | MSN, 0.167 | RASGEF1C, 0.111 | TRIM72, 0.098 |
| CROCCP3, 0.140 | FAS, 0.118 | A_24_P418699, 0.115 | FAM158A, 0.117 | KCNH1, 0.165 | FAM46D, 0.107 | ANKRD19P, 0.098 |
| ASB15, 0.139 | ST5, 0.117 | NETO1, 0.114 | KLHL30, 0.116 | Osta, 0.163 | GUSBP10, 0.105 | PRTFDC1, 0.097 |
| MTMR4, 0.137 | PIP5K1C, 0.117 | COX7A2, 0.113 | CYB5RL, 0.115 | A_24_P219757, 0.158 | P2RY1, 0.104 | C16orf79, 0.097 |
| CNGB3, 0.136 | A_24_P757154, 0.114 | GABARAPL2, 0.112 | A_32_P4294, 0.113 | AKR1C1, 0.157 | MARVELD3, 0.104 | BLMH, 0.093 |
| ANUBL1, 0.133 | AP1M2, 0.113 | RTP1, 0.110 | USP54, 0.113 | S100B, 0.149 | FAIM2, 0.102 | CUST_4410_PI416261804, 0.093 |
| PPP1R16A, 0.129 | A_24_P213134, 0.104 | PANX2, 0.110 | OR4K15, 0.112 | S100A4, 0.144 | LOC728601, 0.102 | A_24_P677734, 0.092 |
| GPR21, 0.128 | SLCO4C1, 0.103 | SVIP, 0.107 | PREX2, 0.111 | PHLDA2, 0.140 | SLC11A2, 0.097 | DARS2, 0.091 |
| COX6A1, 0.127 | EMID2, 0.101 | EHF, 0.105 | SUMO1P1, 0.111 | TCHH, 0.138 | A_24_P942036, 0.096 | A_24_P942630, 0.084 |
| BMX, 0.122 | C6orf142, 0.101 | LCE1B, 0.104 | GFER, 0.109 | A_24_P402090, 0.134 | A_32_P18448, 0.096 | GHDC, 0.084 |

## b

| Visual | Motor | Dorsal Attention | Ventral Attention | Limbic | Control | Default |
|---|---|---|---|---|---|---|
| TNNT2, 0.397* | CARTPT, 0.271* | LOC732327, 0.206* | LIPG, 0.145 | ATP6V0D2, 0.191 | A_24_P204165, 0.204* | SAMD3, 0.131 |
| ASGR2, 0.262* | KLK10, 0.238* | A_24_P178065, 0.153 | A_24_P51184, 0.137 | RRH, 0.155 | BCKDK, 0.147 | FMN1, 0.125 |
| GPX3, 0.240* | ONECUT2, 0.221* | IL1B, 0.149 | CASC5, 0.137 | TRIM54, 0.145 | GUCY2D, 0.143 | VIPR2, 0.122 |
| APOC1, 0.226* | OLR1, 0.209* | SLC1A5, 0.140 | OGFRL1, 0.134 | A_24_P652537, 0.144 | TEX28, 0.137 | ODZ1, 0.119 |
| SOSTDC1, 0.215* | NRGN, 0.154* | A_24_P93184, 0.137 | EXD1, 0.129 | SERPINA1, 0.143 | F9, 0.129 | CSF3R, 0.119 |
| C6orf105, 0.204* | GAS1, 0.152* | SLC27A6, 0.130 | DCAKD, 0.125 | SLC12A5, 0.143 | SPON2, 0.127 | A_24_P76489, 0.117 |
| A_24_P293158, 0.195* | A_32_P65350, 0.152* | FLVCR2, 0.127 | WNT8A, 0.113 | MAP3K10, 0.142 | A_24_P101282, 0.118 | A_24_P372189, 0.115 |
| OVOL2, 0.190* | COL13A1, 0.148* | TREML4, 0.124 | MARCH2, 0.113 | TUBB, 0.141 | GAA, 0.112 | SLC12A2, 0.107 |
| A_24_P292849, 0.185* | ATP1B4, 0.147* | AP003170.1, 0.122 | A_24_P626812, 0.113 | UROS, 0.140 | A_24_P922653, 0.107 | LOC100507191, 0.106 |
| S100A7, 0.170* | MFSD3, 0.147* | SMPD1, 0.120 | RNF152, 0.112 | AACS, 0.139 | DTWD1, 0.104 | SLC17A8, 0.106 |
| PDE6H, 0.160* | VIP, 0.141* | A_32_P111072, 0.119 | HECTD3, 0.108 | MDGA1, 0.135 | EGFL7, 0.101 | AP3B2, 0.105 |
| LOC100131283, 0.158* | MAB21L1, 0.140* | GLI4, 0.109 | PTCH1, 0.105 | C12ORF75, 0.134 | ZNF770, 0.099 | MAFB, 0.103 |
| SLC35F1, 0.157* | KLK5, 0.140* | PLAC2, 0.109 | P2RX6, 0.105 | CD46, 0.133 | KLF5, 0.099 | CALCRL, 0.100 |
| A_23_P256244, 0.155 | HYMAI, 0.14* | A_23_P350059, 0.108 | TBC1D21, 0.105 | SYNCRIP, 0.132 | C8B, 0.098 | A_32_P119165, 0.096 |
| CTXN3, 0.153 | STAC, 0.137* | BIRC7, 0.108 | ANXA6, 0.105 | REXO1, 0.131 | RGS16, 0.097 | DTX3, 0.096 |

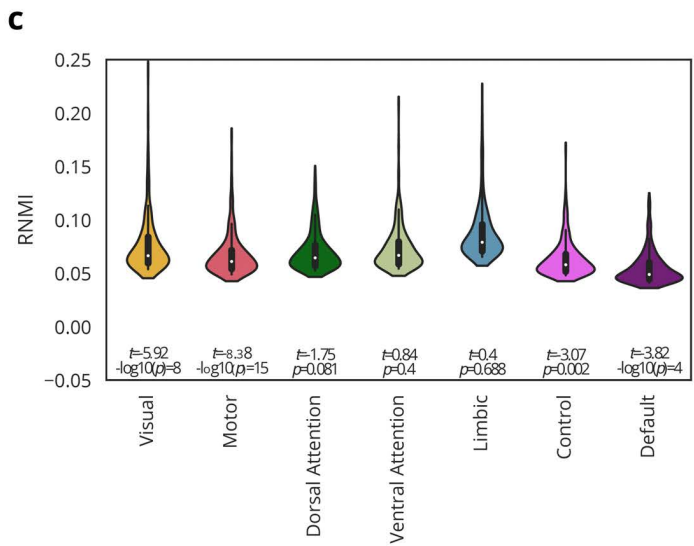
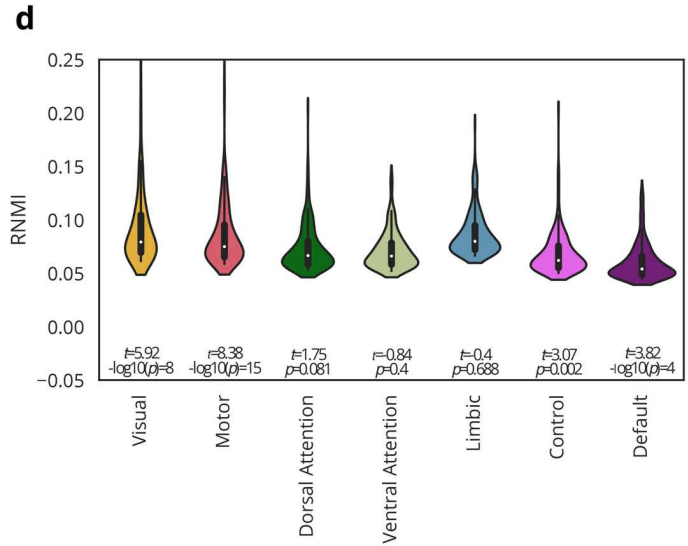

**Extended Data Fig. 13 | PARIS analysis of genes whose variance (SNP) predicts brain connectivity variance. a,b**, PARIS was used to find genes that are differentially useful in predicting connectivity for a particular community versus all other communities. The top 15 genes for (**a**) structural and (**b**) functional connectivity are shown with the FDR corrected p-value for each community. **c,d**, Violin plots of the top 250 genes RNMI values for functional (**c**) and structural (**d**) connectivity. Note that, in both functional and structural connectivity, visual and motor communities have the most genes that strongly differentiate them from the other communities. Also note that in functional connectivity, the motor, visual, dorsal attention, and default mode communities have genes that strongly differentiate communities to a greater extent than structural connectivity does (all communities, $t = 8.99$, $-\log10(p) = 19$, dof=192, individual community comparisons in panels (**c**) and (**d**)), suggesting that genes' SNPs encode functional connectivity in a more modular fashion than they encode structural connectivity.